\documentclass[twocolumn,showpacs,showkeys,amssymb,nobibnotes,superscriptaddress,aps,prd,bibnotes]{revtex4-2}

\usepackage{graphicx}
\usepackage{grffile}
\usepackage{dcolumn}
\usepackage{bm}
\usepackage[T1]{fontenc}
\usepackage{mathptmx}
\usepackage{mathrsfs}
\usepackage{physics}
\usepackage{simplewick}
\usepackage{amsthm,amsmath,amssymb}
\usepackage[bottom]{footmisc}

\usepackage{amssymb}
\usepackage{amsmath}
\usepackage{amsfonts}
\usepackage{bm}
\usepackage{graphicx}
\usepackage{subfigure}
\usepackage[dvipsnames]{xcolor}
\usepackage{calc}
\usepackage{epsfig}
\usepackage{color}
\usepackage[colorlinks,citecolor=blue]{hyperref}

\begin{document}
\bibliographystyle{apsrev4-1}
\newcommand{\be}{\begin{equation}}
\newcommand{\ee}{\end{equation}}
\newcommand{\bs}{\begin{split}}
\newcommand{\es}{\end{split}}
\newcommand{\R}[1]{\textcolor{red}{#1}}
\newcommand{\B}[1]{\textcolor{blue}{#1}}

\title{Tomography of a Macroscopic Quantum State influenced by Classical Self-Gravity}
\author{Wenjie Zhong}
\affiliation{National Gravitational Laboratory, Hubei Key Laboratory of Gravitation and Quantum Physics, School of Physics, Huazhong University of Science and Technology, Wuhan, 430074, China}
\author{Yubao Liu}
\affiliation{State Key Laboratory of Low Dimensional Quantum Physics,
Department of Physics, Tsinghua University, Beijing, 100084, China}
\affiliation{National Gravitational Laboratory, Hubei Key Laboratory of Gravitation and Quantum Physics, School of Physics, Huazhong University of Science and Technology, Wuhan, 430074, China}
\author{Yanbei Chen}
\affiliation{Walter Burke Institute for Theoretical Physics and Theoretical Astrophysics 350-17, California Institute of Technology, Pasadena, CA 91125, USA}
\author{Haixing Miao}
\affiliation{State Key Laboratory of Low Dimensional Quantum Physics,
Department of Physics, Tsinghua University, Beijing, 100084, China}
\author{Yiqiu Ma}
\email{myqphy@hust.edu.cn}
\affiliation{National Gravitational Laboratory, Hubei Key Laboratory of Gravitation and Quantum Physics, School of Physics, Huazhong University of Science and Technology, Wuhan, 430074, China}
\affiliation{Department of Astronomy, School of Physics, Huazhong University of Science and Technology, Wuhan, 430074, China}

\date{\today}

\begin{abstract}
Macroscopic optomechanical systems offer a promising testbed for distinguishing whether gravity acts as a quantum interaction or as a classical field. Schr{\"o}dinger-Newton\,(SN) theory is the nonrelativistic limit of semi-classical gravity where quantum matter couples to classical gravity. Based on SN theory,  this work investigates how classical self-gravity affects continuous quantum state tomography of a macroscopic mechanical oscillator monitored by variable-angle homodyne detection.  In the Schrödinger–Newton (SN) theory, the measurement record arises from a different conditional test mass dynamics from that in quantum-gravity\,(QG)/standard quantum mechanics,  consequently, applying the QG‐optimised reconstruction map introduces an additional state‐dependent contribution. We show that this contribution makes the reconstructed covariance depend on the chosen set of tomography angles and can drive the SN covariance—after QG filtering—outside the standard Gaussian‐covariance domain set by the Heisenberg uncertainty principle. We quantify the resulting QG–SN distinguishability via the Hellinger distance and analyse its dependence on measurement strength and temperature. We then formulate the same issue in the broader setting of nonlinear quantum mechanics: when the system's conditional dynamics during the readout process depends on the state being inferred, the tomographic map acquires nonlinear, model-dependent corrections to the usual Radon or Gaussian reconstruction map. 

\end{abstract}

\maketitle

\section{Introduction}
Semi-classical gravity posits a classical spacetime sourced by the quantum expectation value of the matter density distribution\,\cite{Mueller1962,Rosenfeld1963,Page1981,SalzmanCarlip2006,Carlip_2008,Wald2020,Groshardt2022,Tilloy2016,Tilloy2024,Oppenheim2023A}. Consequently, a test mass's self-gravitating macroscopic quantum state evolves nonlinearly under a state-dependent classical gravitational potential. This dynamics is described by the Schr{\"o}dinger-Newton\,(SN) theory\,\cite{Bahrami_2014,Anastopoulos_2014,Diosi1989,Penrose1996}, the non-relativistic limit of semi-classical gravity. For a macroscopic test mass with mass $M$ and bare pendulum frequency $\omega_m$, the SN theory gives a self-trapping term in the center-of-mass\,(CoM) Hamiltonian\,\cite{Giulini2011,Yang2013,Giulini2014,Grossardt2016,Gan2016Optomechanical},
\be\label{eq:H_sn}
\hat H=\frac{\hat p^2}{2M}+\frac{1}{2}M\omega_m^2\hat x^2+\frac{1}{2}M\omega_{\rm SN}^2(\hat x-\langle \psi|\hat x|\psi\rangle)^2,
\ee
where $\omega_{\rm SN}$ is the SN frequency and $\hat x$ is the center-of-mass displacement operator. The last term is state-dependent: the quantum state evolves in the classical self-gravity potential generated by itself. For an initially prepared squeezed mechanical state with a Gaussian Wigner function, this term makes the second-order correlation function evolve at twice the modified frequency $\omega_q=\sqrt{\omega_m^2+\omega_{\rm SN}^2}$.

In contrast, if gravity has a quantum nature, the CoM Hamiltonian coincides with the standard quantum-mechanical evolution at the bare mechanical frequency $\omega_m$\,\cite{Yang2013,Liu2023}. This is because the leading quantum self-gravity contribution to the Hamiltonian vanishes: for a single rigid test mass, write each atomic position as $\hat{\bm r}_a=\hat{\bm x}+\hat{\bm\rho}_a$, where $\hat{\bm x}$ is the CoM coordinate and $\hat{\bm\rho}_a$ is fixed in the body frame up to internal vibrations. Then the leading internal Newtonian self-energy depends on $|\hat{\bm\rho}_a-\hat{\bm\rho}_b|$, not on $\hat{\bm x}$. It therefore contributes only a constant internal energy and possible renormalizations of fixed mechanical parameters, rather than a state-dependent CoM potential of the form $(\hat x-\langle\hat x\rangle)^2$. In this case, the second-order correlation of an initially squeezed mechanical state will evolve at the frequency $2\omega_m$ rather than $2\omega_q$. Therefore, for the single-test-mass dynamics considered in this paper, the QG prediction for the test-mass Hamiltonian is operationally identical to that of standard quantum mechanics, and we do not distinguish between them in the following discussion. Exploring the different phenomenology predicted by QG and SN theory is part of the broader effort to assess whether gravity must be quantized and how that question may be addressed experimentally\,\cite{Podzien2026,Feng2022,Zhong2025,Bose2017,Marletto2017Gravitationally,Christodoulou2023,krisnanda2020,Belenchia2018,Carney2021Using,Carney_2019,Snowmass2022,Carney2022,Oppenheim2023A,Tilloy2016,Kafri_2014,Miao2020,Datta_2021,Bose2025}. It is important to note that, even if an experiment excludes the SN signatures, this would only rule out the SN description in the present setting; it would not by itself constitute a direct observation of quantum-gravity-mediated effects.

\begin{figure}[h]
\centering
\includegraphics[width=0.45\textwidth]{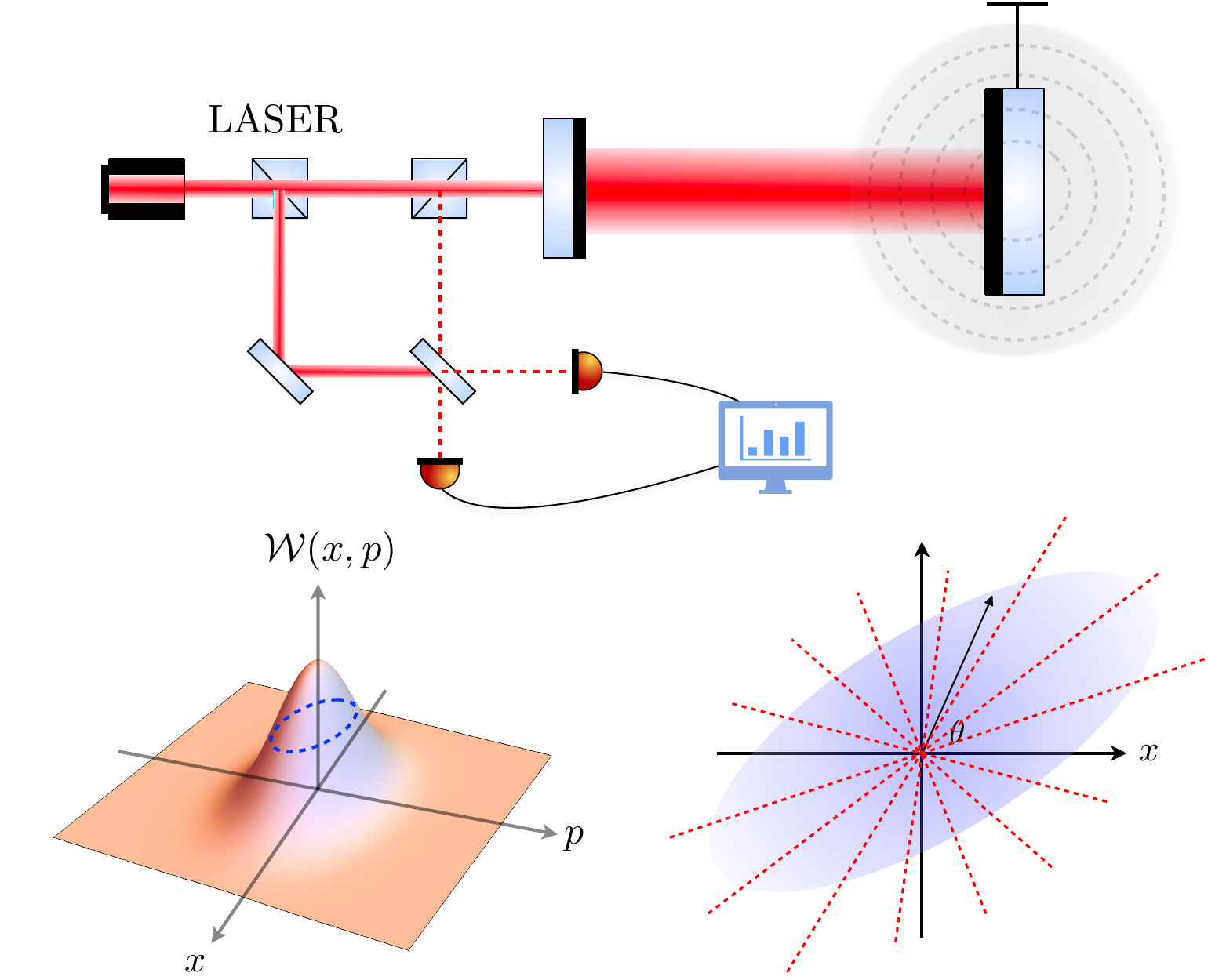}
\caption{Concept of tomography for a self-gravitating mechanical oscillator. The displacement of a suspended mirror is continuously encoded in the outgoing optical field and read out with variable-angle homodyne detection. The resulting measurement records provide noisy projections of the mechanical phase-space distribution.}
\label{fig:Wigner_function}
\end{figure}


A mechanical quantum state can be probed by coupling the test mass to an optical field, during which information about the test mass flows into the optical field, as illustrated in Fig.~\ref{fig:Wigner_function}. Properly filtering the optical measurement data allows us to extract the information of the test mass quadrature and hence reconstruct the test mass quantum state, a procedure known as \emph{quantum state tomography}\,\cite{Miao2010,Chen_2013,Wieczork2015}. During the optomechanical interaction, entanglement between the test mass and the light field is continuously established. Upon measurement of the output light, this joint optomechanical quantum state is projected onto the Hilbert space of the test mass motion, generating a conditional quantum state $|\psi_c\rangle$ of the test mass. This process occurs continuously in a continuous measurement setup and is closely related to quantum-limited optomechanical state preparation, trajectory tracking, and macroscopic measurement back action\,\cite{Braginsky1995,Aspelmyer2014,Rossi2019Observing,Cripe2019,Delic2020}. Consequently, the influence of the measurement manifests itself in replacing $\langle \psi|\hat x|\psi\rangle$ in Eq.\,\eqref{eq:H_sn} by the conditional mean $x_c = \langle \psi_c|\hat x|\psi_c\rangle$---which is the Causal Conditional Formalism of Schr{\"o}dinger-Newton\,(CCSN) theory\,\cite{Liu2023,Liu2024,Miki2025}. The magnitude of the SN contribution in the CCSN theory can then be estimated as $\sim M\omega_{\rm SN}^2\langle\psi_c|(\hat x- x_c)^2|\psi_c\rangle = M \omega_{\rm SN}^2 V_{xx}^c$, where $V_{xx}^c$ is the conditional variance of the test mass motion. In an ideal case with sufficiently strong optical power (i.e., strong measurement strength and negligible shot noise), the conditional variance tends to zero because we achieve complete and noiseless extraction of the test mass information. In this strong measurement limit, the SN contribution is negligible, and the tomography of the test mass quantum state proceeds as in standard quantum mechanics.

However, as we will show in this work, when the optical power is not strong enough to sufficiently suppress the SN contribution, the SN term cannot be naively neglected when probing a self-gravitating quantum state within semi-classical gravity. This implies that quantum gravity and semi-classical gravity generate different conditional dynamics during the tomography process. The main task of this work is therefore operational and hypothesis-driven: we investigate how an SN-generated measurement record is reconstructed when processed with the standard QG/standard-quantum tomography map, and identify signatures of the resulting model mismatch.

In this work, we deliberately adopt the QG/standard-quantum tomography filter as the common operational reconstruction map. Starting from the same initially prepared Gaussian squeezed state $|\psi(0)\rangle$, we ask how the reconstructed covariance changes if the measurement record is generated by QG dynamics or by SN dynamics. This is not a claim that the QG filter is the optimal SN filter. Rather, it is a controlled way to expose model mismatch: the same data-analysis prescription is applied to two different underlying dynamical laws, and the resulting reconstructed states $\tilde{\rho}^{\rm QG}_{\rm QG/SN}(T)$ are compared.

Our main findings are as follows. First, the SN correction to the tomographic error is controlled by the conditional covariance trajectory and is therefore quadrature-angle dependent. As a result, different choices of tomographic angles can lead to different SN-reconstructed uncertainty ellipses under the same fixed QG reconstruction map, whereas the QG reconstruction is angle-set independent. Second, the SN correction is not a positive noise covariance. When the QG reconstruction map is applied to an SN-generated record, the determinant of the inferred covariance matrix can fall below the standard Heisenberg bound or even become nonpositive, which can be interpreted as a model-mismatch signature of SN dynamics compared to that of QG. Third, the mismatch between the QG and SN reconstructions depends differently on the measurement strength and temperature, which we quantify using the Hellinger distance. Finally, we show that this fixed-map mismatch problem is a specific instance of a more general obstruction in tomography of nonlinear quantum dynamics: the measurement process can change the state variables that also determine the subsequent dynamical map.

The structure of this paper is organized as follows. Section~\ref{sec:continuous_tomography} develops the basic theory of quantum state tomography in an optomechanical system where the test mass is influenced by classical self-gravity. Section~\ref{sec:reconstruction} investigates how the Schr{\"o}dinger-Newton effect modifies tomography performed with the standard quantum-mechanical filter. Section~\ref{sec:experiment} analyzes possible experimental signatures that distinguish SN theory from the QG reference case using a fixed operational reconstruction map. Section~\ref{sec:nonlinear_tomography} formulates the corresponding issue for tomography in general nonlinear quantum mechanics, and Sec.~\ref{sec:conclusion} combines this viewpoint with the SN unknown-state problem and concludes the paper.

\section{Quantum State Tomography}\label{sec:continuous_tomography}
The purpose of this section is to identify what is actually measured in a continuous optomechanical tomography experiment, and how the answer is changed by the Schr{\"o}dinger-Newton (SN) self-gravity term when the data are processed with the standard QG/standard-quantum filter. In ordinary quantum mechanics, or in the QG reference model used in this paper, the measured variance consists of the desired mechanical quadrature variance plus the familiar tomographic noise from optical shot noise, radiation-pressure back action, and thermal force noise. In the SN case, the measurement record is generated by a different conditional covariance dynamics. Applying the same filter to the same nominal quadrature therefore produces an additional contribution that is attributed to the state-dependent SN self-gravity. The formulae below make this separation explicit, and they reduce continuously to the QG result presented in\,\cite{Miao2010,Chen_2013} when $\omega_{\rm SN}\to0$. Different from\,\cite{Miao2010,Chen_2013} where they work in the Heisenberg picture, we present a different derivation in the Schr{\"o}dinger picture, which turns out to be more convenient than that in Heisenberg picture in the SN theory where the conditional mean $x_c=\langle\psi_c|\hat x|\psi_c\rangle$ is involved in the Hamiltonian.

\subsection{State Tomography: Problem Setup}
Tomography is performed by letting the optical field carry away information about the mechanical displacement and then post-processing the classical photocurrent, which is generated by a homodyne detection of the output optical quadratures. The optomechanical interaction and the coupling between the intra-cavity field and the external optical field are described by
\be
\hat H_{\rm om}=-\hbar\alpha(\hat a+\hat a^\dag)\hat x,\quad\hat H_{\rm ext}=i\hbar\sqrt{2\gamma}(\hat a^\dag\hat a_{\rm in}-h.c),
\ee
where $\alpha=\sqrt{2P_c\omega_0/\hbar L c}$ is the linear optomechanical interaction strength, $P_c$ is the intra-cavity power, $L$ is the cavity length, $\gamma$ is the cavity bandwidth, and $\omega_0$ is the optical pumping frequency. The corresponding optomechanical measurement frequency is defined as $\Omega_q=\sqrt{\hbar\alpha^2/M}$. This frequency sets the natural scale for the tomography filters: increasing $\Omega_q$ means extracting mechanical information faster, but also increasing the radiation-pressure disturbance that must be canceled by an appropriate homodyne angle.

Starting from the optomechanical Hamiltonian above, we describe the continuous readout in the Schr{\"o}dinger picture using the stochastic-master-equation, or quantum-trajectory, formalism reviewed in Ref.~\cite{Chen_2013} and detailed derived for SN theory in~\cite{Liu2023}.  Each weak homodyne measurement gives an outcome carrying partial information about the mechanical position and updates the mechanical state to a conditional state $\ket{\psi_c(t)}$; repeating this measurement update in the continuous limit gives a stochastic conditional trajectory. If the measured optical quadrature is chosen by the time-dependent homodyne angle $\phi(t)$, the corresponding classical photocurrent is therefore written in terms of the conditional mean $x_c(t)=\langle\psi_c(t)|\hat x|\psi_c(t)\rangle$, with the optical vacuum fluctuation appearing as the Wiener process:

\be\label{eq:measurement_data}
\tilde{y}(t)=\alpha x_c(t)\sin\phi(t)+\frac{dZ(t)}{\sqrt{2}dt},
\ee
where $dZ(t)$ denotes the Wiener increment associated with the measurement process. A filter window $W(t)$ compresses the continuous record into a single measured quantity over the measurement duration $T_{\rm obs}$,
\be\label{eq:Y_data}
Y=\int_{0}^{T_{\rm obs}}dt\ W(t)\tilde{y}(t).
\ee

Having compressed the continuous photocurrent into the filtered outcome $Y$, we now interpret its variance in the context of state tomography. Since the measurement record is Gaussian and the filtering is linear, $Y$ is a one-dimensional Gaussian random variable, fully characterized by its variance. We introduce a Gaussian decomposition of the filtered data as $Y = Y_s + Y_n$, where $Y_s$ denotes the component whose variance exactly matches that of the initial mechanical quadrature, and $Y_n$ accounts for the erroneous deviation. Operationally, as we shall prove, the part of $\sigma^2_{YY}$ proportional to the initial quadrature covariance identifies $Y_s$:
\begin{equation}
    \sigma^2_{Y_s Y_s} \propto \left\langle\psi(0)\left|\left(\hat{x}\cos\theta+\frac{\hat{p}}{M\Omega_q}\sin\theta\right)^2\right|\psi(0)\right\rangle,
\end{equation}
where $\sigma^2_{AB}\equiv\overline{AB}-\overline{A}\,\overline{B}$, and the overbar denotes ensemble averaging over noise realizations. All remaining erroneous variance is grouped into the residual error $\sigma^2_e$.

Our goal is to minimize $\sigma^2_e$ to faithfully reconstruct the initial quantum state. This translates into an optimization problem---both physical and mathematical---of choosing the temporal window $W(t)$ and homodyne angle $\phi(t)$ so that the optical record preserves the desired quadrature information while optimally rejecting imprecision, back-action, and thermal noise. Such a process has been thoroughly analyzed for standard quantum mechanics in\,\cite{Miao2010}, in the following, we will examine how the SN self-gravity term affects this tomographic process.

\subsection{Optomechanical Dynamics during Tomography}
In describing how the conditional mechanical states evolve during the optomechanical quantum tomography process, we need two pieces of dynamics. First, the conditional mean follows a stochastic trajectory driven by the measurement process and by the thermal Langevin force. Second, the conditional covariance evolves deterministically according to a Riccati equation; it determines both the strength of the measurement update and the size of the SN self-gravity correction. This is the point at which the QG and SN descriptions separate: the conditional mean propagates with the bare mechanical response characterized by $\omega_m$, while the conditional covariance feels the SN-modified oscillator frequency $\omega_q$.

We first define the response matrices used below. Let $\nu=m,q$ denote the bare mechanical response and the SN-modified response, respectively. The corresponding drift matrices are
\begin{equation}\label{eq:drift_matrices}
    \bm{A}_\nu =
    \begin{pmatrix}
        0 & 1/M \\
        -M\omega_\nu^2 & -\gamma_m
    \end{pmatrix},
    \qquad \nu\in\{m,q\},
\end{equation}
where $\gamma_m$ is the mechanical damping rate. We also define the damping-modified mechanical frequencies $\omega_{\nu c}=\sqrt{\omega_\nu^2-\gamma_m^2/4}$ and write the response functions as
\begin{equation}\label{eq:response_matrix}
  e^{\bm{A}_\nu t} =
  \begin{pmatrix}
    f_1^\nu(t) & G_\nu(t) \\[6pt]
    M\dot f_1^\nu(t) & M\dot G_\nu(t)
  \end{pmatrix},
  \qquad
  f_2^\nu(t)\equiv M\Omega_qG_\nu(t).
\end{equation}
Explicitly,
\begin{align}\label{eq:response_functions}
  f_1^\nu(t) &=
  e^{-\frac{\gamma_m}{2}t}
  \left[
  \cos\omega_{\nu c}t
  +\frac{\gamma_m}{2\omega_{\nu c}}\sin\omega_{\nu c}t
  \right],\notag\\
  f_2^\nu(t) &=
  \frac{\Omega_q}{\omega_{\nu c}}
  e^{-\frac{\gamma_m}{2}t}\sin\omega_{\nu c}t.
\end{align}
These two functions $f_{1}(t)$ and $f_{2}(t)$ correspond to the free evolution of the two mechanical quadratures $x(0)$ and $p(0)$, respectively.

\emph{Conditional mean}\,--- Let the conditional displacement and momentum be collected into the vector
\begin{equation}
    \bm{r}_c(t)= [x_c(t), p_c(t)]^T,
\end{equation}
whose stochastic equation of motion is
\begin{equation}\label{eq:conditional_mean_sde_matrix}
    d\bm{r}_c(t)=\bm{A}_m\bm{r}_c(t)\,dt
    +\bm{\Gamma}(t)\,dZ(t)
    +\bm{\Xi}_{\rm th}\,dZ_n(t),
\end{equation}
where $dZ_n(t)$ is a Wiener increment for the thermal bath coupling and is independent of the measurement-driven $dZ(t)$. The deterministic drift in this equation is $\bm{A}_m$, because the SN force is centered on the conditional mean and therefore does not shift the mean trajectory itself. The SN effect instead enters through the conditional covariance appearing in the measurement update below.
The measurement-noise vector is
\begin{equation}\label{eq:Gamma_definition}
    \bm{\Gamma}(t)
    =\sqrt{2}\alpha
    \begin{pmatrix}
        \sin\phi(t)\,V_{xx}^c(t)\\
        \sin\phi(t)\,V_{xp}^c(t)+\dfrac{\hbar}{2}\cos\phi(t)
    \end{pmatrix}.
\end{equation}
The two components of $\bm{\Gamma}$ have a direct interpretation. The term proportional to $V_{xx}^c$ is the Bayesian update of the inferred position, while the term proportional to $V_{xp}^c$ and $\hbar\cos\phi/2$ describes the associated update of momentum, including the part tied to optical back action. The classical thermal noise drive in Eq.~\eqref{eq:conditional_mean_sde_matrix} is specified by
\begin{equation}\label{eq:thermal_mean_noise_vector}
    \bm{\Xi}_{\rm th}=[0,   \sqrt{2M\gamma_m k_B T}]^T
\end{equation}
Here $T$ is the thermal bath temperature and $k_B$ is Boltzmann's constant, and the above thermal vector is derived based on the correlation function of classical thermal Langevin force below:
\begin{equation}
    \overline{f_{\rm th}(t)f_{\rm th}(t')}
    =2M\gamma_m k_B T\,\delta(t-t'),
\end{equation}
where we take the high temperature limit $k_BT/\hbar\omega_m\gg1$. Note that, as shown in\,\cite{Helou2017,Liu2026}, there exist two different prescriptions in treating thermal noise in SN theory. In this work, we take the classical thermal noise prescription where the thermal Langevin force is treated as a classical force noise affecting the quantum trajectory of the test mass, while the conditional covariance of the test mass is only affected by the quantum radiation pressure noise. Related quantum-classical and measurement-feedback viewpoints have also been discussed in Refs.~\cite{Tilloy2016,Diosi1998}.
 
Integrating Eq.\,\ref{eq:conditional_mean_sde_matrix} with initial condition of the mean displacement and momentum $\bm{r}_c(0)=(x_0,p_0)^T$ leads to:
\begin{align}\label{eq:conditional_mean_matrix_solution}
    \bm{r}_c(t)
    =&\,e^{\bm{A}_m t}\bm{r}_c(0)
    +\int_0^t e^{\bm{A}_m(t-s)}
    \bm{\Gamma}(s)\,dZ(s)\notag\\
    &+\int_0^t e^{\bm{A}_m(t-s)}
    \bm{\Xi}_{\rm th}\,dZ_n(s),
\end{align}
The conditional displacement is therefore the sum of three physically distinct contributions,
\begin{equation}\label{eq:evolution of x}
    x_c(t) = x^{(0)}(t)+x^m(t)+x^{\rm th}(t).
\end{equation}
The three terms are
\begin{align}
    x^{(0)}(t) =& x_0f_1^m(t)+\frac{p_0}{M\Omega_q}f_2^m(t), \label{eq:free_evolution_x}\\
    x^m(t) =& \int^t_0 dZ(s)\, \bm{H}^T(t-s)\bm{\Gamma}(s), \label{eq:measurement of x}\\
    x^{\rm th}(t) =& \sqrt{\frac{2\gamma_m k_B T}{M\Omega_q^2}} \int^t_0 dZ_n(s)\, f_2^m(t-s).\label{eq:measurement of th}
\end{align}
The $\bm{H}^T(t)$ is defined as:
\begin{equation}\label{eq:H_response_row}
    \bm{H}^T(t)=
    \begin{pmatrix}
        f_1^m(t), & \dfrac{f_2^m(t)}{M\Omega_q}
    \end{pmatrix},
\end{equation}
where $f_1^m(t)$ and $f_2^m(t)$ are defined in Eq.\,\eqref{eq:response_functions}.
Here $x^{(0)}(t)$ denotes the free evolution of the initial first moments,
with $x_0$ and $p_0$ representing the quantum expectation values of the
initial position and momentum, respectively; $x^m(t)$ is the stochastic
displacement inferred from and driven by the optical measurement; and
$x^{\rm th}(t)$ is the response to the classical thermal force noise. This decomposition is useful because only the first term contains the desired initial quadrature signal, while the latter two terms become added tomographic noise after the measurement record is filtered.

\emph{Conditional covariance matrix}\,---
The conditional covariance matrix is defined as:
\begin{equation}
    \bm{V}^c(t) = \begin{pmatrix} V_{xx}^c(t) & V_{xp}^c(t) \\   V_{xp}^c(t) & V_{pp}^c(t) \end{pmatrix},
\end{equation}
and satisfy the Riccati equation:
\begin{equation}\label{eq:Riccati_matrix}
    \dot{\bm{V}}^c = \bm{A}_q \bm{V}^c + \bm{V}^c \bm{A}_q^T + \bm{D}_0 - \bm{\Gamma}\bm{\Gamma}^T,
\end{equation}
where $\bm{D}_0 = \operatorname{diag}(0, \hbar^2\alpha^2/2)$. The SN modification appears through $\bm{A}_q$, so the conditional covariance oscillates at the doubled wave-packet-breathing frequency $2\omega_q$. The diffusion matrix $\bm{D}_0$ is contributed by radiation-pressure back-action force noise, which tends to increase the momentum variance. The nonlinear term $-\bm{\Gamma}\bm{\Gamma}^T$ is the information gain from the measurement; it narrows the conditional state because $\bm{\Gamma}$ is proportional to the current conditional uncertainty. The competition between these two effects produces a transient oscillatory covariance and, for a fixed measurement strategy, a steady conditional covariance when $\dot{\bm{V}}^c=0$. 

\subsection{Tomography: signal variance and residual error}
We now evaluate the variance of the filtered measurement outcome, $\sigma^2_{YY}$, directly from the stochastic measurement record. This calculation identifies a boundary term proportional to the initial mechanical covariance, which we define as the tomographic signal variance $\sigma^2_{Y_sY_s}$, and groups all remaining terms into the residual tomographic error $\sigma^2_e$.  Substituting Eq.\,\eqref{eq:measurement_data} into Eq.\,\eqref{eq:Y_data}, and defining
\begin{equation}
    C_{xdZ}(t,t')\equiv
    \overline{x_c(t) dZ(t')/dt'},
\end{equation}
and the filter function $g_1(t)\equiv W(t)\cos\phi(t)$, $g_2(t)\equiv W(t)\sin\phi(t)$, we obtain:
\be
\begin{split}
    \overline{Y^2}=&\int_{0}^{T_{\rm obs}}dt\int_{0}^{T_{\rm obs}}dt'\left[\alpha^2 g_2(t)g_2(t')\overline{x_c(t)x_c(t')}\right.\\
    &\left.+\sqrt{2}\alpha g_2(t)W(t')C_{xdZ}(t,t')+\frac{1}{2}W(t)W(t')\delta(t-t')\right].\label{eq:E[YY]}
\end{split}
\ee

The first term is the autocorrelation of the conditional mechanical trajectory. Using Eq.~\eqref{eq:evolution of x}, it decomposes into
\begin{equation}
\overline{x_c(t)x_c(t')} = x^{(0)}(t)x^{(0)}(t') + \overline{x^m(t)x^m(t')} + \overline{x^{\rm th}(t)x^{\rm th}(t')}.
\end{equation}
Since $x^{(0)}(t)$ arises from the free evolution of the initial-state quantum expectation values, it is a deterministic $c$-number. Hence, the contribution $x^{(0)}(t)x^{(0)}(t')$ is exactly canceled by the square of the mean, $\overline{Y}^2$. The two statistically independent noise-driven paths $x^m(t)$ and $x^{\rm th}(t)$ contribute the tomography error. In addition, the measurement-induced part also contains a boundary term proportional to the initial covariance; this is the term from which tomography extracts the information of the initial state.

For instance, using $\overline{f_{\rm th}(t)f_{\rm th}(t')}=2M\gamma_m k_B T\,\delta(t-t')$, we can derive the auto-correlation of the thermal-noise displacement $x^{\rm th}(t)$ as:
\begin{align}
  \overline{x^{\rm th}(t)x^{\rm th}(t')}=\int_{0}^{\tau}ds\,
  \bm{H}^T(t-s)\bm{D}_{\rm th}\bm{H}(t'-s),
\end{align}
where $\tau=\min(t,t')$. The thermal diffusion matrix $\bm{D}_{\rm th}$ is given by:
\begin{equation}
  \bm{D}_{\rm th}=
  \begin{pmatrix}
    0&0\\
    0&\hbar M\Omega_{\rm th}^2
  \end{pmatrix},
  \qquad
  \Omega_{\rm th}^2 \equiv \frac{2\gamma_m k_B T}{\hbar},
\end{equation}
which describes the diffusion of the quantum trajectory driven by the classical thermal noise. The $\Omega_{\rm th}$ is the characteristic frequency for the thermal noise.

For the measurement-induced term, the autocorrelation function is
\be
\overline{x^m(t)x^m(t')} = \int_0^{\tau} ds\, \bm{H}^T(t-s) \bm{\Gamma}(s) \bm{\Gamma}^T(s) \bm{H}(t'-s),
\ee 
and the information about the initial state can be separated by using the Riccati equation to replace $\bm{\Gamma}(s) \bm{\Gamma}^T(s)$,
\be
\bm{\Gamma}\bm{\Gamma}^T = \bm{A}_q \bm{V}^c + \bm{V}^c \bm{A}_q^T + \bm{D}_0 - \dot{\bm{V}}^c,
\ee
The time derivative term is the useful part of this manipulation because it can be integrated by parts:
\begin{align}\label{eq:integral_by_parts}
    &\overline{x^m(t)x^m(t')}\supset -\int_0^{\tau} ds\, \bm{H}^T(t-s) \dot{\bm{V}}^c(s) \bm{H}(t'-s) \notag\\
    =& -\Big[ \bm{H}^T(t-s) \bm{V}^c(s) \bm{H}(t'-s) \Big]_{s=0}^{s=\tau} \notag\\
    &- \int_0^{\tau} ds\, \bm{H}^T(t-s) \left[ \bm{A}_m \bm{V}^c(s) + \bm{V}^c(s) \bm{A}_m^T \right] \bm{H}(t'-s),
\end{align}
where we have performed the integration by parts and used the equality $d\bm{H}(t)/dt = \bm{A}_m^T \bm{H}(t)$.

Using the symmetry $\bm{V}^{cT} = \bm{V}^c$ and the identity $\bm{H}(0) = (1, 0)^T$, the boundary term at $s=\tau$ simplifies to $\bm{H}^T(|t-t'|) \bm{V}^c(\tau) \bm{H}(0)$. The boundary term at $s=0$ contains $\bm{V}^c(0)$, which will contribute to the desired initial-covariance contribution. The back-action autocorrelation can therefore be written as
\be\label{eq:xm_xm_final}
\begin{split}
    \overline{x^m(t)x^m(t')} =& \bm{H}^T(t)\bm{V}^c(0)\bm{H}(t') - \bm{H}^T(|t-t'|) \bm{V}^c(\tau) \bm{H}(0) \\
    &+ \int_0^{\tau} ds\, \bm{H}^T(t-s) \bm{M}_c(s) \bm{H}(t'-s),
\end{split}
\ee
where  $\bm{M}_c(s) $ is defined as 
\be
\bm{M}_c(s) \equiv \bm{A}_{\rm SN}  \bm{V}^c(s) + \bm{V}^c(s) \bm{A}_{\rm SN} ^T + \bm{D}_0.
\ee
and $\bm{A}_{\rm SN}$ is
\begin{equation}
    \bm{A}_{\rm SN} \equiv \bm{A}_q - \bm{A}_m =
    \begin{pmatrix}
        0 & 0 \\
        -M\omega_{\rm SN}^2 & 0
    \end{pmatrix}.
\end{equation}
As we shall see later that this $\bm A_{\rm SN}$ matrix contributed to the SN effect on the tomography error.

Combining $\overline{x^m(t)x^m(t')}$ and $\overline{x^{\rm th}(t)x^{\rm th}(t')}$ gives the conditional-trajectory contribution to $\overline{Y^2}$:
\begin{align}\label{eq:xc_autocorrelation_contribution}
&\alpha^2\int_{0}^{T_{\rm obs}}dt
  \int_{0}^{T_{\rm obs}}dt'\,
  g_2(t)g_2(t')\notag\\
  &\times\Bigg[
    \bm{H}^T(t)\bm{V}^c(0)\bm{H}(t')
    -\bm{H}^T(|t-t'|)\bm{V}^c(\tau)\bm{H}(0)\notag\\
  &\qquad
    +\int_0^\tau ds\,\bm{H}^T(t-s)
    \left[\bm{M}_c(s)+\bm{D}_{\rm th}\right]
    \bm{H}(t'-s)
  \Bigg].
\end{align}

The second term in Eq.~\eqref{eq:E[YY]} is the correlation between optical shot noise and the measurement-induced displacement. Since only $x^m$ is driven by $dZ(t')$, we have
\begin{equation}
    C_{xdZ}(t,t')=
    \bm{H}^T(t-t')\bm{\Gamma}(t')\Theta(t-t'),
\end{equation}
where $\Theta$ is the Heaviside's step function. 
Substituting into Eq.\,\eqref{eq:E[YY]} leads to:
\begin{equation}
\begin{split}\label{eq:cross_correlation_integral}
  &\sqrt{2}\alpha \int_{0}^{T_{\rm obs}}dt\int_{0}^{T_{\rm obs}}dt'[g_2(t)W(t')C_{xdZ}(t,t') ] \\
  =& \alpha^2\int_{0}^{T_{\rm obs}}dt\int_{0}^{T_{\rm obs}}dt' g_2(t)g_2(t') \Big[\bm{H}^T(|t-t'|) \bm{V}^c(\tau) \bm{H}(0)\Big] \\
  &+ \Omega_q\int_{0}^{T_{\rm obs}}dt\int_{t}^{T_{\rm obs}}dt' g_1(t)g_2(t') f_2^m(t'-t).
\end{split}
\end{equation}
Note that the $\alpha^2$ term in Eq.\,\eqref{eq:cross_correlation_integral} exactly cancels the same term in Eq.\,\eqref{eq:xc_autocorrelation_contribution}. 

Finally, the last term in Eq.~\eqref{eq:E[YY]}, $W(t)W(t')\delta(t-t')/2$, is the tomography error contributed by optical shot noise. Integrating the delta function and using $g^2_1(t) + g^2_2(t)=W^2(t)$, we obtain
\be
\overline{Y^2}_{\rm sh}=\frac{1}{2}\int_0^{T_{\rm obs}} dt [g^2_1(t) + g^2_2(t)]=\frac{1}{2}\int^{T_{\rm obs}}_0W^2(t)
\ee

\emph{Signal variance}\,--- We can now write the decomposition $\sigma^2_{YY}=\sigma^2_{Y_sY_s}+\sigma^2_{e}$. To make the filtered record estimate the mechanical quadrature at angle $\theta$, we impose
\begin{equation}\label{eq:filter_constraint}
    \int_0^{T_{\rm obs}} dt\, g_2(t)f_1^m(t) = \cos\theta, \quad \int_0^{T_{\rm obs}} dt\, g_2(t)f_2^m(t) = \sin\theta,
\end{equation}
which fixes the signal response to the initial position and momentum. The initial-covariance contribution is then
\begin{align}\label{eq:signal_raw}
    \sigma^2_{Y_sY_s} =& \alpha^2\int_{0}^{T_{\rm obs}}dt\int_{0}^{T_{\rm obs}}dt' g_2(t)g_2(t')\bm{H}^T(t)\bm{V}^c(0)\bm{H}(t') \notag\\
     =& \frac{\Omega_q}{2}\left[\cos^2\theta \frac{V_{xx}^c(0)}{\delta x_q^2} + \sin2\theta \frac{V_{xp}^c(0)}{\delta x_q\delta p_q} + \sin^2\theta \frac{V_{pp}^c(0)}{\delta p_q^2}\right],
\end{align}
where $\delta x_q=\sqrt{\hbar/(2M\Omega_q)}$ and $\delta p_q=\sqrt{\hbar M\Omega_q/2}$. Apart from the overall factor $\Omega_q/2$, this is exactly the variance of the dimensionless mechanical quadrature of the initial state that is to be probed:
\begin{equation}
     \sigma^2_{Y_sY_s}=\frac{\Omega_q}{2}  \langle\psi(0)| \hat X^2_\theta|\psi(0)\rangle,\quad \hat X_\theta=\frac{\hat x(0)}{\delta x_q}\cos\theta
    +\frac{\hat p(0)}{\delta p_q}\sin\theta.
\end{equation}

\emph{Tomography error}\,---  The residual tomography error functional $\sigma^2_{e}[g_1,g_2]$ in the SN case decomposes as:
\be
    \sigma^2_{e}[g_1,g_2] = \sigma^2_{\rm QG}[g_1,g_2]+ \sigma^2_{\rm SN}[g_1,g_2],
\ee
where the first term is the tomography error in the QG, or equivalently standard quantum-mechanical, reference case:
\be\label{eq:sigma2_QG}
\begin{split}
 \sigma^2_{\rm QG}= \frac{1}{2}\int_{0}^{T_{\rm obs}}dt & W^2(t)+\Omega_q\int_{0}^{T_{\rm obs}}dt\, g_1(t)j_2(t) \\
    &+ \left(\Omega_{\rm th}^2+\frac{\Omega_q^2}{2}\right)\int_{0}^{T_{\rm obs}}dt\ j_2(t)^2,
\end{split}
\ee
where we define:
\be\label{eq:j12}
\begin{split}
j_{1,2}(t) = \int_t^{T_{\rm obs}} dt' g_2(t')f_{1,2}^m(t'-t),
 \end{split}
\ee
and $j_{1,2}(t)$ are related by the following equality:
\be\label{eq:j12_relation}
\frac{dj_2(t)}{dt}+\Omega_q j_1(t)= \gamma_m j_2(t).
\ee
The three terms in Eq.\,\eqref{eq:sigma2_QG} are, respectively, optical shot noise, the correlation between the quantum trajectory and shot noise, and the combined thermal-force and radiation-pressure force noise. The role of $g_1$ is especially transparent here: by choosing the measured optical quadrature appropriately, one can make the second term cancel the radiation-pressure part of the third term in the optimized QG filter.

The second component, $\sigma^2_{\rm SN}[g_1, g_2]$, comes directly from the difference between the SN and bare drift matrices, $\bm{A}_{\rm SN}=\bm{A}_q-\bm{A}_m$:
\begin{align}\label{eq:N_SN_functional}
    \sigma^2_{\rm SN}[g_1, g_2]= -\omega_{\rm SN}^2\int_{0}^{T_{\rm obs}}dt\, j_2(t)\left[h_{xx}(t) j_1(t)+  h_{xp}(t) j_2(t)\right],
\end{align}
where, for notational simplicity, we have defined the dimensionless conditional covariance as:
\begin{equation}
  h_{xx}(t)\equiv\frac{V_{xx}^c(t)}{\delta x_q^2},\quad
  h_{xp}(t)\equiv\frac{V_{xp}^c(t)}{\delta x_q\delta p_q},\quad
  h_{pp}(t)\equiv\frac{V_{pp}^c(t)}{\delta p_q^2}.
\end{equation}

This SN contribution vanishes when $\omega_{\rm SN}=0$. Unlike ordinary added noise, however, it depends on the full evolution history of $V_{xx}^c(t)$ and $V_{xp}^c(t)$. This feature will lead to two important consequences: (1) Since those conditional covariances are themselves shaped by the homodyne angle, measuring different quadratures $\hat X_\theta$ can generate different covariance histories. This angle-dependent history is the key reason that the SN correction cannot, in general, be absorbed into a single angle-independent uncertainty ellipse in the same way as the QG tomography error.  (2) The $\sigma^2_{\rm SN}$ could ``distort" the signal $ \sigma^2_{Y_s Y_s}$ since the evolution of the conditional covariance contains the information of $V_{xx}^c(0),V_{xp}^c(0)$ due to the non-stationary nature of the tomography process. Although we can not completely separate the $V_{xx}^c(0),V_{xp}^c(0)$ from the $\sigma^2_{\rm SN}$, there is a revised version which could make the distortion more clear: substituting the equality Eq.\,\eqref{eq:j12_relation} into Eq.\,\eqref{eq:N_SN_functional}, we obtain:
\be
\begin{split}
\sigma_{\rm SN}^2& \approx
 -\omega_{\rm SN}^2 \left[\frac{h_{xx}(0)\sin^2\theta}{2\Omega_q} \right.\\
&\left.+ \int_0^{T_{\rm obs}}dt\,j^2_2(t) \left( 2h_{xp}(t)- \frac12\sin^2\phi(t)\,h^2_{xx}(t) \right) \right],
\end{split}
\ee
in deriving which we need to perform the integral by parts on $dj_2(t)/dt$ and use the boundary conditions $j_2(T_{\rm obs})=0$, $j_2(0)=\sin\theta$. Although the second term still carries the information of $V_{xx}^c(0),V_{xp}^c(0)$, the first term, for instance at $\theta=\pi/2$, clearly reduces the contribution of $V^c_{pp}(0)$ to the signal term $\sigma^2_{Y_s Y_s}$. In a certain parameter region, such a distortion could be very strong as we shall show in the next section.

Last comment on this section is about the methodology: why do we prefer the Schrödinger picture to the Heisenberg picture above? As discussed in\,\cite{Miki2025}, the Heisenberg-picture method used for QG tomography can be generalized formally to SN/CCSN by treating the conditional mean as a causal classical feedback signal. In steady-state settings such as the experimental protocol analyzed in\,\cite{Helou2017,Liu2023,Liu2024,Miki2025,Liu2026,Yan2025}, which target distinguishing SN/QG by measuring the steady spectrum of the outgoing light or related semiclassical-gravity signatures, this leads to a Wiener-filter description which can be applied to calculate the required optical spectrum. However, tomography is a finite-time, nonstationary inference problem: the conditional covariance trajectory entering the SN feedback depends on the initial covariance itself. Therefore, the Heisenberg-picture approach does not restore the QG-like signal-plus-independent-noise decomposition.

\section{Effect of the SN Term on State Tomography}\label{sec:reconstruction}
As mentioned in the Introduction section, the main task of this work is to investigate how an SN-generated measurement record is reconstructed under the standard QG/standard-quantum tomography procedure. State tomography in the QG reference model requires us to optimize the filter functions $g_1(t)$ and $g_2(t)$ (or equivalently $W(t)$ and $\phi(t)$) to minimize the error functional. We therefore first derive the QG filter and the corresponding QG reconstruction map, and then study the deviations that arise when the same reconstruction map is applied to data generated by Schr{\"o}dinger-Newton dynamics rather than QG dynamics.

\subsection{System parameters}
In the following discussion, we will numerically analyze the tomography process of a sample optomechanical system, with parameters listed in Table~\ref{tab:system_parameters}. 
\begin{table}[h]
    \centering
    \begin{tabular}{|c|c|c|}
    \hline
    \textbf{Parameter} & \textbf{Symbol} & \textbf{Value} \\\hline
    Cavity length & $L$ & $2$ m\\
    Mirror mass & $M$ & $0.2$ kg \\
    Mechanical quality factor & $Q$  & $10^7$ \\
    Mirror eigenfrequency &$\omega_m/(2\pi)$ & $4\times10^{-3}$ Hz\\
    SN frequency & $\omega_{\rm SN}/(2\pi)$ & $7.8\times10^{-2}$ Hz\\
    Optical wavelength & $\lambda$ & $1064$ nm\\
    Finesse & $\mathcal{F}_{\rm cav}$ & $300$\\
    Initial state & $\bm{V}_0$ & squeezed Gaussian, see text\\
        \hline
    \end{tabular}
    \caption{Key parameters of the optomechanical system with a single movable mirror under SN self-gravity.}
    \label{tab:system_parameters}
\end{table}
Here, we chose osmium as the bulk material of the movable mirror. The bare mechanical resonant frequency $\omega_m$ is set at milli-Hertz since a larger $\omega_m$ will reduce the relative contribution of the SN effect. At the same time, the mean thermal occupation number $n_{\rm th}=k_BT/\hbar\omega_mQ$ can not be too high, otherwise the thermal noise will dominate the tomography error and reduce the distinguishability of the reconstructed state under the QG and SN theory. Therefore, the requirement of a pronounced SN effect during the tomography process in a thermal environment leads to a stringent mechanical quality factor. While this paper is about the theoretical principles, we admit that such a mechanical quality factor for a milli-Hertz macroscopic pendulum oscillator and the cryogenic environment challenge the experimental feasibility. Relevant experimental platforms and low-frequency optomechanical technologies have been actively developed in recent years\,\cite{Westpal2021,Smetana2024,Matsumoto2020}.

The quadratic SN Hamiltonian in Eq.~\eqref{eq:H_sn} is valid when the spatial uncertainty of the crystal CoM is much smaller than the zero-point fluctuation of the ions around their lattice sites $\Delta x_{\rm c.m.}\ll \Delta x_{\rm zp}$. The atomic zero-point displacement in an osmium crystal is estimated as $\Delta x_{\rm zp}\sim 2\times10^{-12}\,{\rm m}$. The initial mechanical covariance is denoted by $\bm{V}_0\equiv\bm{V}^c(0)$. In the numerical examples, we use a Gaussian squeezed state characterized by
\begin{equation}
    V_{0,xx}=0.2\,\frac{\hbar}{2M\omega_q},\,
    V_{0,xp}=\frac{\hbar}{2},\,
    V_{0,pp}=10\,\frac{\hbar M\omega_q}{2}.
\end{equation}
The corresponding position uncertainty is
\begin{equation}
    \Delta x_{\rm c.m.}(0)=\sqrt{V_{0,xx}}=\sqrt{0.2\hbar/2M\omega_q},
\end{equation}
which is $\sim 10^{-17}\,{\rm m}$ for the parameters in Table~\ref{tab:system_parameters}. The quadratic potential condition $\Delta x_{\rm c.m.}/\Delta x_{\rm zp}\lesssim 10^{-4}$ is therefore satisfied.

\subsection{Optimal filter and state reconstruction in QG}
To extract the quadrature information at a certain tomography angle $\theta$ with maximum precision, we seek the filter functions $g_1(t)$ and $g_2(t)$ that minimize $\sigma^2_{\rm QG}[g_1, g_2]$ subject to the constraints Eq.\,\eqref{eq:filter_constraint}. This is the standard linear optomechanical filtering problem: the optical readout carries imprecision noise, radiation pressure produces back action, and the homodyne angle can be chosen so that the back-action contribution is canceled in the estimator. 
Closely related optimal-filter and state-estimation problems have been studied in standard quantum optomechanics~\cite{Miao2010,Chen_2013,Wieczork2015,Wiener1964Extrapolation}.  Using the method of Lagrange multipliers, we construct the functional:
\begin{equation}
  \begin{split}
    \mathcal{L}[g_1,g_2]=&\sigma^2_{\rm QG}[g_1, g_2]-\mu_1\left[\int_0^{T_{\rm obs}} dt\, g_2(t)f_1^m(t)-\cos\theta\right]\\
    &-\mu_2\left[\int_0^{T_{\rm obs}} dt\, g_2(t)f_2^m(t)-\sin\theta\right],
  \end{split}
\end{equation}
where $\mu_1$ and $\mu_2$ are the Lagrange multipliers. 

Functional derivation with respect to $g_1(t),g_2(t)$ leads to the optimization condition of the tomography filters. For instance, variation with respect to $g_1(t)$ leads to one optimization condition:
\begin{align}
  \frac{\delta \mathcal{L}}{\delta g_1(t)}= g_1(t) + \Omega_q j_2(t) = 0. \label{eq:g1_solution}
\end{align}
The physical interpretation of this condition is that the homodyne angle $\phi(t)$ is carefully chosen so that the quantum optomechanical back-action is completely evaded in the readout channel, which is widely-discussed variational readout method in improving the sensitivity of a laser interferometer gravitational wave detector\,\cite{Kimble2001,Miao2010,Vyatchanin1996,Regal2017}. Substituting this relation back to the noise functional leads to 
\begin{align}\label{eq:optimal_noise}
  \sigma^2_{\rm QG}[g_2] =& \frac{1}{2}\int_{0}^{T_{\rm obs}}dt\ g^2_2(t) + \Omega_{\rm th}^2\int_{0}^{T_{\rm obs}}dt\ j^2_2(t).
\end{align}
where the $\Omega_q$ term vanishes---a signature that the quantum measurement back-action is completely evaded.

In addition, variation with respect to $g_2(t)$ yields an integral equation
\begin{align}\label{eq:g2_solution}
  \frac{\delta \mathcal{L}}{\delta g_2(t)}
  =&\,
  g_2(t)
  +2\Omega_{\rm th}^2
  \int_{0}^{T_{\rm obs}}dt'\,g_2(t')
  \int_{0}^{\min\{t,t'\}}d\tau \notag\\
  &\times
  f_2^m(t-\tau)f_2^m(t'-\tau)
  -\mu_1f_1^m(t)-\mu_2f_2^m(t)=0,
\end{align}
which is a Fredholm integral equation of the second
kind on the finite time interval $[0,T_{\rm obs}]$.  In general, this equation does
not admit a simple closed-form solution for arbitrary finite $T_{\rm obs}$.
However, in the long observational time approximation,
Eq.~\eqref{eq:g2_solution} can be solved and such an asymptotic solution for $g_2(t)$
has a characteristic evolution frequency
\begin{equation}
\begin{split}
  \Omega_v
  &=
  \frac{1}{2}
  \sqrt{
    -\gamma_m^2
    +2\omega_m^2
    +2\sqrt{\omega_m^4+2\Omega_{\rm th}^2\Omega_q^2}
  }\\
  &\approx 2^{-1/4}\sqrt{\Omega_{\rm th}\Omega_q}.
  \end{split}
  \label{eq:Omega_v}
\end{equation}
For the finite-time estimate used below, we choose the observation time to be
one characteristic period of this asymptotic optimal filter, $ T_{\rm obs}=2\pi/\Omega_v$.
The solutions to Eqs.~\eqref{eq:g1_solution} and \eqref{eq:g2_solution}
therefore provide the theoretically optimal filter functions
$g_1^{\rm QG}(t)$ and $g_2^{\rm QG}(t)$, with the finite-time window fixed by
the characteristic frequency of the analytic long-time solution:
\begin{align}
  g_2^{{\rm opt},X}(t)&\simeq 2\Omega_v e^{-\Omega_v t}\cos\Omega_v t,\\
  g_2^{{\rm opt},P}(t)&\simeq
  2\sqrt{2}\frac{\Omega_v^2}{\Omega_q}
  e^{-\Omega_v t}\sin\left(\Omega_v t-\frac{\pi}{4}\right).
\end{align}
The corresponding $g_1$ filter follows from
Eq.~\eqref{eq:g1_solution} and decomposes in the same way,
$g_1^{\rm opt}(t;\theta)=g_1^{{\rm opt},X}(t)\cos\theta+g_1^{{\rm opt},P}(t)\sin\theta$, with
\begin{align}
  g_1^{{\rm opt},X}(t)&\simeq
  \frac{\Omega_q^2}{\Omega_v}e^{-\Omega_v t}\sin\Omega_v t,\\
  g_1^{{\rm opt},P}(t)&\simeq
  -\sqrt{2}\Omega_q e^{-\Omega_v t}
  \sin\left(\Omega_v t+\frac{\pi}{4}\right).
\end{align}
Note that these analytical results are a very good approximation to the exact optimal filter solved numerically, and the exact numerical filter will be used in the following analysis.
It is also important to note that the two optimisation conditions Eq.\,\eqref{eq:g1_solution} and Eq.\,\eqref{eq:g2_solution} are linear, therefore we can construct the optimal filter for an arbitrary tomographic angle as:
\begin{equation}\label{eq:g_2_decompose}
  g_2^{\rm QG}(t;\theta)= g_2^{X,\rm QG}(t)\cos\theta+g_2^{P,\rm QG}(t)\sin\theta,
\end{equation}
where $g_2^{X,\rm QG}$ and $g_2^{P,\rm QG}$ are the two optimal filters associated with the $X$- and $P$-quadrature constraints. As expected, these equations reduce to the optimal continuous
measurement filters derived in the Heisenberg picture under standard quantum optomechanics~\cite{Miao2010}.

Since the initial state is a Gaussian state with an elliptic cross-section of the Wigner function,  we need to obtain the uncertainties of three different quadratures $\Theta\equiv\{\theta_1,\theta_2,\theta_3\}$ to reconstruct the initial state. Defining the uncertainty vector as $\bm{\sigma}_{YY}^2 = (\sigma^2_{YY}(\theta_1), \sigma^2_{YY}(\theta_2), \sigma^2_{YY}(\theta_3))^T$, we have
\begin{equation}
    \bm{\sigma}_{YY}^2 = \alpha^2 \bm{\Lambda}(\Theta) \bm{v}(0) + \bm{\sigma}^2_{\rm QG} + \bm{\sigma}^2_{\rm SN},
\end{equation}
where
\begin{equation}
\begin{split}
       & \bm{\Lambda}(\Theta) = [
        \bm{\lambda}(\theta_1),\,\,
        \bm{\lambda}(\theta_2),\,\,
        \bm{\lambda}(\theta_3)]^T\\
    & \bm{\lambda}(\theta)=\left(\cos^2\theta,\,\frac{\sin2\theta}{M\Omega_q},\,\frac{\sin^2\theta}{M^2\Omega_q^2}\right).
\end{split}
\end{equation}
and
$\bm{v}(0) \equiv (V_{xx}^c(0), V_{xp}^c(0), V_{pp}^c(0))^T$. The $\bm{\sigma}^2_{\rm QG}$ and $\bm{\sigma}^2_{\rm SN}$ are column vectors encapsulating the corresponding QG noise and SN correction terms evaluated at the three selected angles.

With three distinct angles\,($\theta_i \neq \theta_j \pm n\pi$), the projection matrix $\bm{\Lambda}$ is non-singular. By applying the inverse operator $\bm{X} \equiv (\alpha^2 \bm{\Lambda})^{-1}$, we extract the reconstructed covariance vector $\bm{v}^{\rm recon}$:
\begin{align}
    \bm{v}^{\rm recon} &= \bm{X} \bm{\sigma}_{YY}^2 \notag \\
    &= \bm{v}(0) + \bm{X}\bm{\sigma}^2_{\rm QG} +\bm{X}\bm{\sigma}^2_{\rm SN} . \label{eq:reconstructed_v_vector}
\end{align}
The corresponding covariance matrix is obtained by reshaping this three-component vector,
\begin{equation}
  \bm{V}^{\rm recon}=\mathcal{V}[\bm{v}^{\rm recon}],\quad {\rm where}\quad
  \mathcal{V}[\bm{a}] \equiv
  \begin{pmatrix}
    a_1 & a_2\\
    a_2 & a_3
  \end{pmatrix}.
\end{equation}

\subsection{Schr{\"o}dinger-Newton  effect}
When the same QG reconstruction procedure is applied to Schr{\"o}dinger-Newton generated data, the above QG tomography structure is challenged. Substituting the filter function derived in the previous subsection into Eq.\,\ref{eq:optimal_noise}, we obtain the QG-reference tomography error for the $\hat X_\theta$ quadrature as
\be\label{eq:sigma_QG}
\sigma_{\rm QG}^2(\theta)=\bm{\Pi}^T(\theta)\bm{\sigma}_{\rm QG,ref}^2
\ee
where
\be
\bm{\Pi}(\theta)=
 \begin{pmatrix}
\cos^2\theta-\sin\theta\cos\theta\\
 \sin^2\theta-\sin\theta\cos\theta\\
 \sin2\theta
  \end{pmatrix},\,\,
\bm{\sigma}_{\rm QG,ref}^2\equiv
    \begin{pmatrix}
    \sigma_{\rm QG}^2(0)\\
    \sigma_{\rm QG}^2(\pi/2)\\
    \sigma_{\rm QG}^2(\pi/4)
  \end{pmatrix},
\ee 
and $\bm{\sigma}_{\rm QG,ref}^2$ is the tomography error at three reference quadrature angles.
Substituting the above equation into Eq.\,\eqref{eq:reconstructed_v_vector} where we ignore the SN-term  in the QG case, we obtain:
\be
\bm{v}^{\rm recon} = \bm{v}(0) + \bm{X}(\Theta)\bm{\Pi}^T(\Theta)\bm{\sigma}_{\rm QG,ref}^2.
\ee
Note that direct calculation can prove that the coefficient matrix $\bm{X}(\Theta)\bm{\Pi}^T(\Theta)$ is $\theta-$independent:
\be
\bm{X}(\Theta)\bm{\Pi}^T(\Theta)=
\begin{pmatrix}
1&0&0\\
-M\Omega_q/2& -M\Omega_q/2&M\Omega_q\\
0&(M\Omega_q)^2&0
\end{pmatrix},
\ee
which means that the tomography error is independent of the reference tomography angles. For example, instead of $(0,\pi/4,\pi/2)$, we may choose $(0,\pi/6,\pi/3)$ or other admissible angles, and the same tomography uncertainty ellipse will be obtained.

\begin{figure}
  \centering
  \includegraphics[width=1\columnwidth]{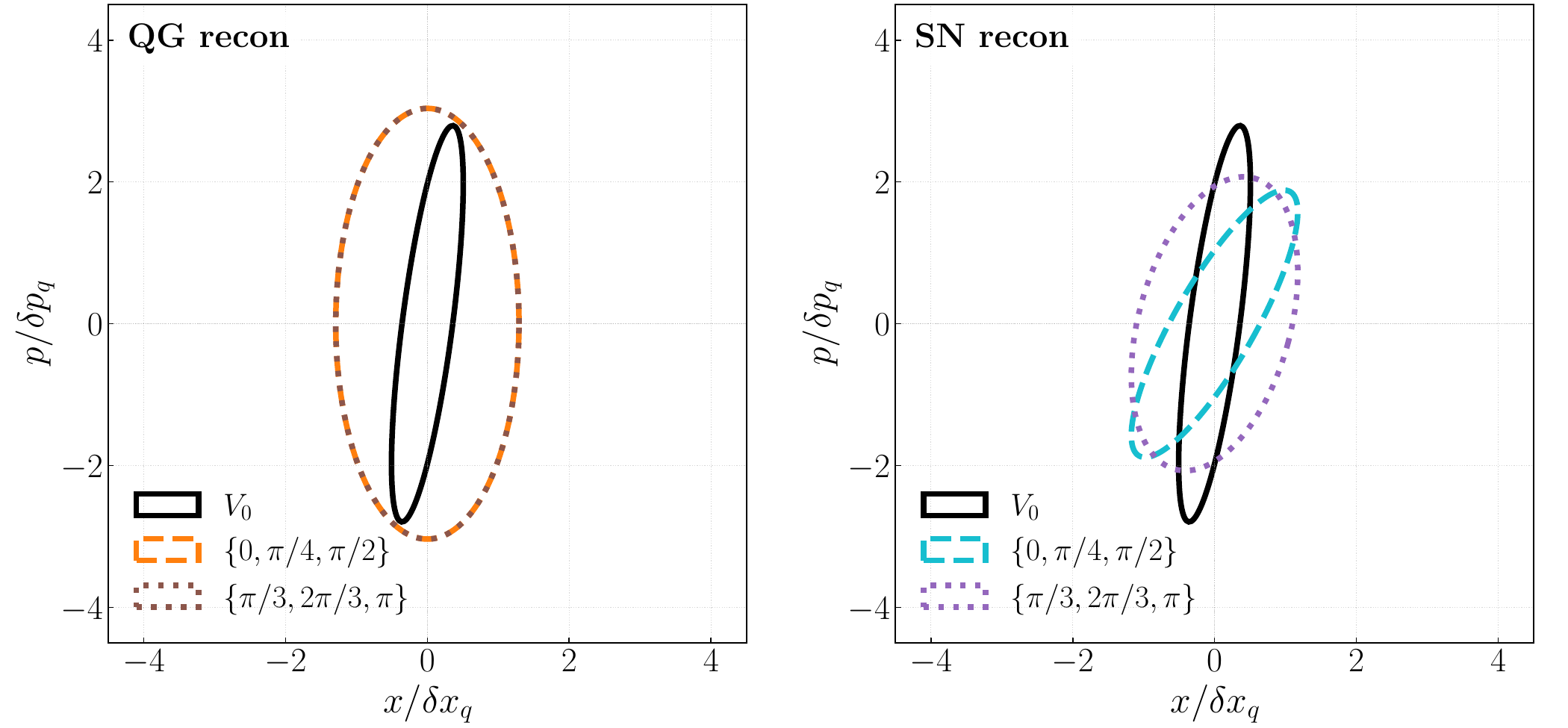}
  \caption{Dependence of the reconstructed covariance on the choice of tomographic angles. In the QG reconstruction (left), the two angle sets $\Theta_A=\{0,\pi/4,\pi/2\}$ and $\Theta_B=\{\pi/3,2\pi/3,\pi\}$ lead to the same uncertainty ellipse. In the SN reconstruction (right), the correction to the tomographic noise depends on the conditional covariance trajectory associated with each homodyne angle, so the same two angle sets produce different uncertainty ellipses. The black ellipse denotes the initial covariance $\bm{V}_0$; here $\Omega_q/2\pi=0.1$ Hz and $T=0.3$ mK.}
  \label{fig:SN_angle_sets}
\end{figure}

However, the above property of QG tomography is no longer preserved when the same reconstruction map is applied to SN-generated data. This is because the SN correction $\sigma^2_{\rm SN}[g_1,g_2]$ in Eq.~\eqref{eq:N_SN_functional} depends on the evolution history of $V_{xx}^c(t)$ and $V_{xp}^c(t)$, which themselves depend on the tomography angle $\theta$ as discussed in the last subsection. We denote such dependence as $[V_{xx}^c(t|\theta),V_{xp}^c(t|\theta),V_{pp}^c(t|\theta)]$. In detail, when deriving Eq.\,\eqref{eq:sigma_QG} for an SN-generated record, we obtain
\begin{equation}
\begin{split}
  \sigma^2_{\rm SN}(\theta)
  =-\omega_{\rm SN}^2[A_{\rm SN}(\theta)\cos^2\theta
  +B_{\rm SN}(\theta)\sin^2\theta\\
  +2C_{\rm SN}(\theta)\sin\theta\cos\theta],
  \end{split}
\end{equation}
where
\be
\begin{split}
A_{\rm SN}&(\theta)=\int^{T_{\rm obs}}_0dt j_2^X(t)\left[j_1^X(t)h_{xx}(t|\theta)+j_2^X(t)h_{xp}(t|\theta)\right],\\
B_{\rm SN}&(\theta)=\int^{T_{\rm obs}}_0dt j_2^P(t)\left[j_1^P(t)h_{xx}(t|\theta) +j_2^P(t)h_{xp}(t|\theta)\right],\\
C_{\rm SN}&(\theta)=\\
&\int^{T_{\rm obs}}_0dt \left[j_2^X(t)j_2^P(t)h_{xp}(t|\theta)+j_1^{\{ X}(t)j_2^{P\}}(t)h_{xx}(t|\theta)\right],
\end{split}
\ee
where $j^{X/P}_{1/2}$ are defined by substituting Eq.\,\eqref{eq:g_2_decompose} into Eq.\,\eqref{eq:j12}:
\be
j^\theta_{1/2}(t)\equiv j^{X}_{1/2}\cos\theta+j^{P}_{1/2}\sin\theta,
\ee
and $j_1^{\{ X}j_2^{P\}}\equiv j_1^{X}j_2^{P}+j_1^{P}j_2^{X}$.
It is this $\theta$ dependence of the coefficients $A_{\rm SN}$, $B_{\rm SN}$ and $C_{\rm SN}$ that prevents us from
writing the tomography error in the form of Eq.\,\eqref{eq:sigma_QG}. Hence, the same fixed QG reconstruction map does not generate a unique uncertainty ellipse when applied to different angle sets $(\theta_1,\theta_2,\theta_3)$ for SN-generated data. Figure~\ref{fig:SN_angle_sets} illustrates this effect with two choices of tomographic angles: $\Theta_A=\{0,\pi/4,\pi/2\}$ and $\Theta_B=\{\pi/3,2\pi/3,\pi\}$. The QG-filtered SN reconstructions yield visibly different uncertainty ellipses.

If an SN-generated record is reconstructed using the QG tomographic map, the inferred matrix may leave the standard Gaussian-covariance domain. This behavior arises because the SN-induced contribution to the tomography error does not necessarily act as added positive noise; depending on the filter and conditional trajectory, it can also reduce the covariance inferred by the mismatched reconstruction model.

This is in sharp contrast to the QG case with the reconstructed matrix
 $\bm{V}_{\rm QG}^{\rm recon}=\bm{V}_0+\bm{V}_{\rm QG}^{\rm error}$ where $\bm{V}_{\rm QG}^{\rm error}\equiv \mathcal{V}[\bm{X}\bm{\sigma}^2_{\rm QG}]$. Note that the $\bm{V}_{\rm QG}^{\rm error}$ is positive semidefinite, $\bm{V}_{\rm QG}^{\rm error}\succeq 0$: i.e. for an arbitrary vector $\bm{x}=r(\cos\theta,\sin\theta)^{T}$, the following quadratic form is always positive
\begin{equation}
\bm{x}^{T}\bm{V}_{\rm QG}^{\rm error}\bm{x}=r^{2}\sigma_{\rm QG}^{2}(\theta)>0,
\end{equation}
which can be seen by using Eq.~\eqref{eq:optimal_noise}. 
The determinant of the reconstructed covariance matrix is:
\be
\begin{split}
\det\bm{V}_{\rm QG}^{\rm recon}&=\det\left(\bm{V}_0+\bm{V}_{\rm QG}^{\rm error}\right)\\
&=\det\left(\bm{I}+\bm{V}_0^{-1/2}\bm{V}_{\rm QG}^{\rm error}\bm{V}_0^{-1/2}\right)\cdot\det\bm{V}_0\\
&=(1+\lambda_1)(1+\lambda_2)\cdot\det\bm{V}_0\geq\det\bm{V}_0\geq\frac{\hbar^{2}}{4},
\end{split}
\ee
where we have used the fact that,  for $\bm{V}_0\succ 0$ and $\bm{V}^{\rm error}_{\rm QG}\succeq 0$,  the matrix $\bm{V}_0^{-1/2}\bm{V}_{\rm QG}^{\rm error}\bm{V}_0^{-1/2}$ is positive semidefinite, with eigenvalues $\lambda_1,\,\lambda_2\ge0$.
For any initial state, $\det\bm{V}_0\geq\hbar^{2}/4$, and therefore
\begin{equation}
\det\bm{V}_{\rm QG}^{\rm recon}\geq\frac{\hbar^{2}}{4}.
\end{equation}
Thus, the QG reconstruction of a physical Gaussian initial state remains inside the standard Gaussian-covariance domain.

\begin{figure}
  \centering
  \includegraphics[width=\columnwidth]{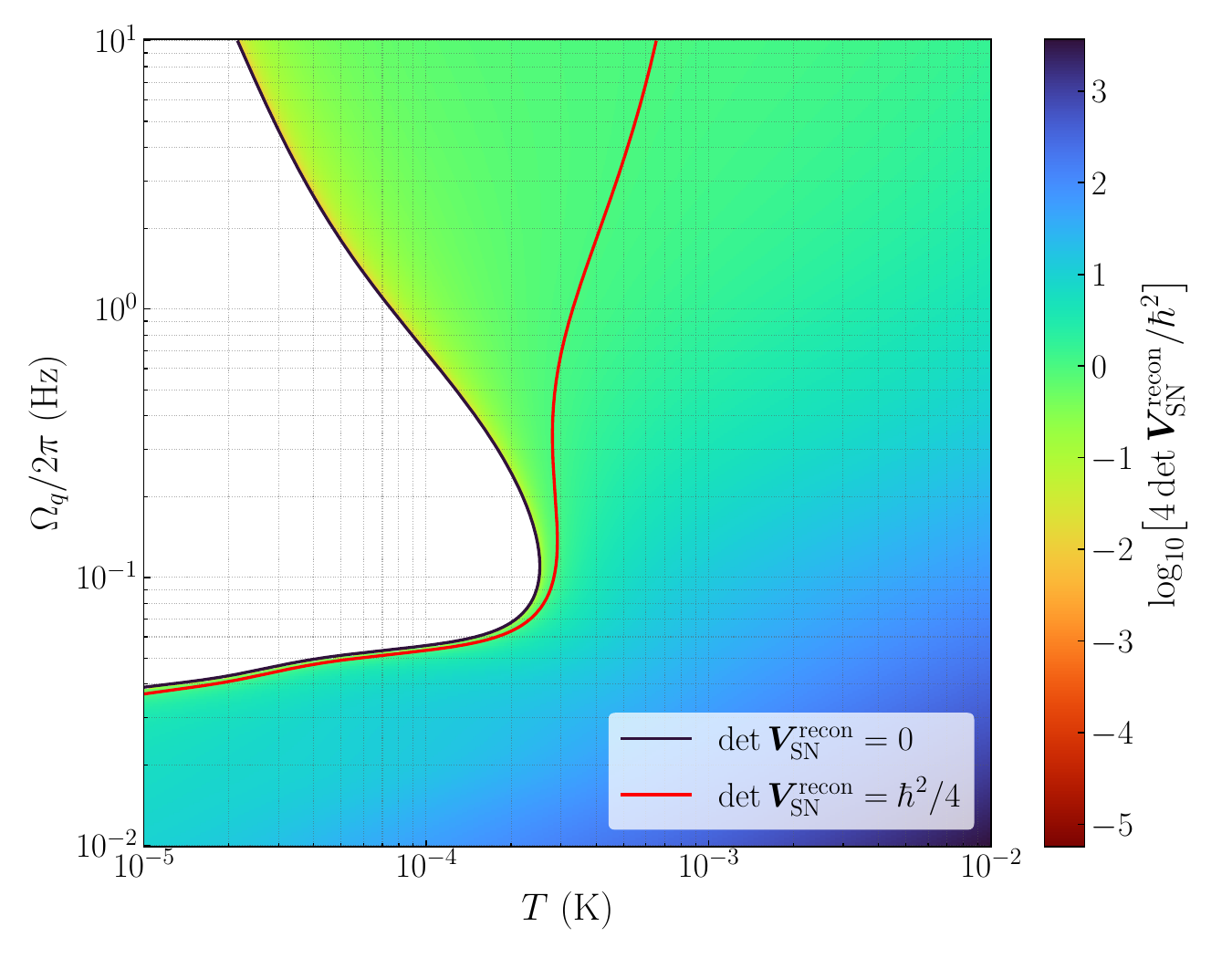}
  \caption{Gaussian-covariance consistency of the SN reconstruction obtained with the QG tomographic map. The color scale shows the magnitude of $\det\bm{V}_{\rm SN}^{\rm recon}$. The dark contour marks $\det\bm{V}_{\rm SN}^{\rm recon}=0$, while the cyan contour marks the standard Gaussian determinant boundary $\det\bm{V}_{\rm SN}^{\rm recon}=\hbar^2/4$. Crossing these contours identifies parameter regions where the QG-filtered SN covariance is either nonpositive or below the usual Gaussian covariance bound.}
  \label{fig:SN_uncertainty_violation}
\end{figure}

By contrast, $\bm{V}_{\rm SN}^{\rm error}=\mathcal{V}[\bm{X}(\bm{\sigma}^2_{\rm QG}+\bm{\sigma}^2_{\rm SN})]$ is not always positive semidefinite, and hence $\bm{V}_{\rm SN}^{\rm recon}\succeq\bm{V}_0$ no longer holds. The mismatched QG reconstruction can therefore return an inferred SN covariance with determinant below $\hbar^{2}/4$, or even with negative determinant, when the result is interpreted as an ordinary Gaussian covariance.  Moreover, we scan over different parameter settings in Fig.~\ref{fig:SN_uncertainty_violation}, where the initial covariance matrix is reconstructed from tomography at $\Theta=(0, \pi/4, \pi/2)$. The color bar corresponds to the value ${\rm log}_{10}[4{\rm det}(\bm{V}_{\rm SN}^{\rm recon})/\hbar^2]$, and the two contours mark $\det(\bm{V}_{\rm SN}^{\rm recon})=\hbar^2/4$ and $\det(\bm{V}_{\rm SN}^{\rm recon})=0$. A closely related below-Heisenberg-bound feature also appears in the output-light-spectrum analysis of Ref.~\cite{Liu2024}, where the optical spectrum obtained from the SN/CCSN dynamics can cross the standard quantum-mechanical Heisenberg spectral bound under the usual interpretation. In both cases, the point is that SN dynamics modifies the relation between the measured observable and the standard quantum covariance/noise.

\begin{figure*}
    \centering
    \subfigure{
    \includegraphics[width=\textwidth]{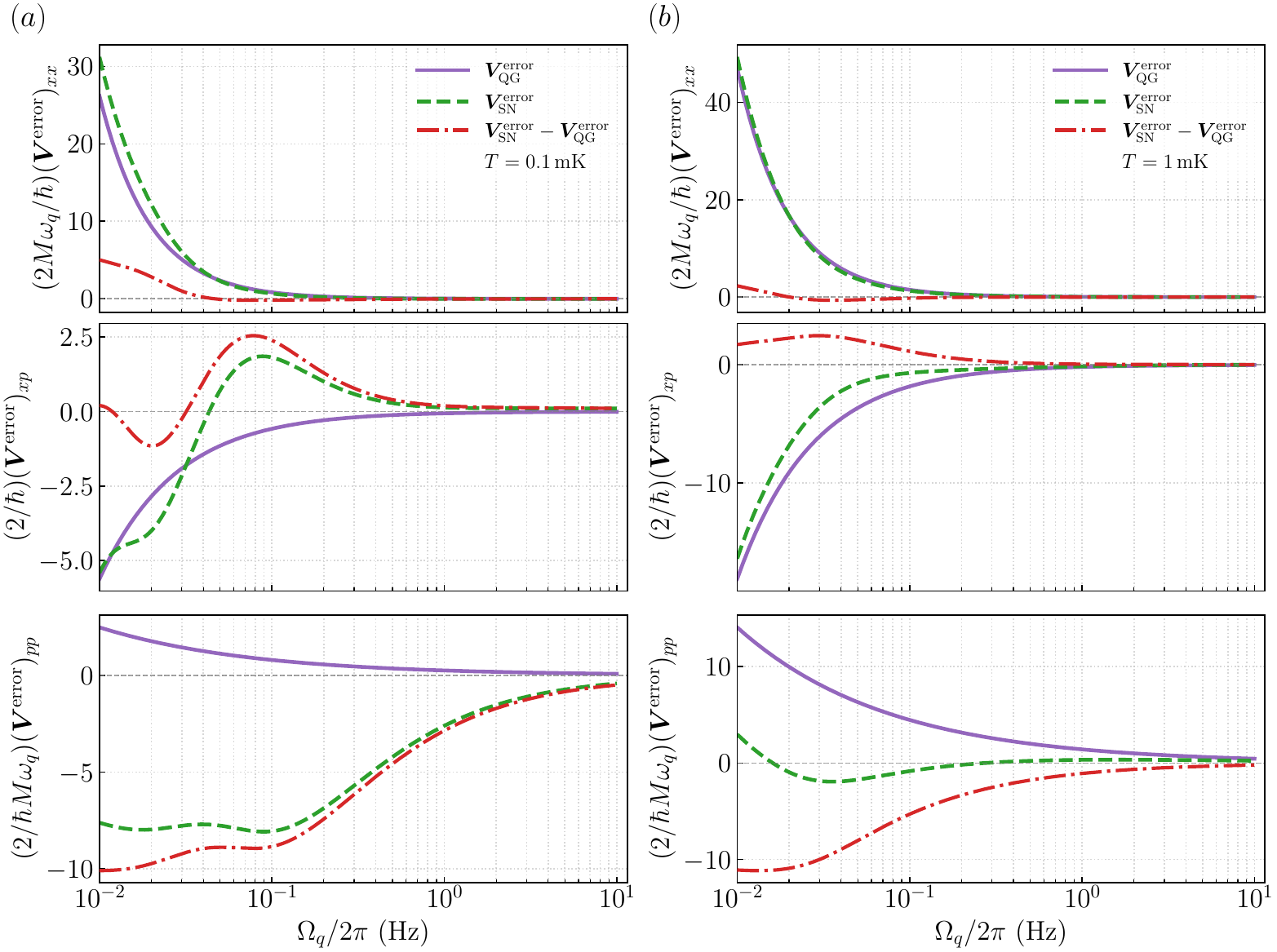}
    \label{fig:error_components}}
    \caption{Tomographic error-matrix components at two different temperatures. Panel (a) shows the low-temperature $T=0.1$\,mK case: the three panels show the $xx$, $xp$, and $pp$ components as functions of the measurement strength $\Omega_q/2\pi$. Purple solid curves give the QG error, green dashed curves give the SN error, and red dot-dashed curves give the SN-induced correction $\bm{V}^{\rm error}_{\rm SN}-\bm{V}^{\rm error}_{\rm QG}$. At low temperature this correction is comparable to the QG tomographic noise over part of the scan and can drive the $xp$ and $pp$ components negative. Panel (b) shows the relatively high-temperature $T=1$\,mK case with the same notation and line styles. Compared with the low-temperature case, thermal force noise increases the QG reconstruction error while leaving the conditional-covariance-driven SN correction nearly unchanged, so the relative SN contribution is reduced.}   \label{fig:error_components_temperature}
\end{figure*}

In Fig.\,\ref{fig:error_components_temperature}, we present the tomography error matrix components $[\bm{V}^{\rm error}_{\rm QG/SN}]_{ij}$, with $i,j=x,p$, at two different temperatures; these components correspond to the tomography uncertainty at mechanical quadrature angles $\Theta=(0,\pi/4,\pi/2)$. 
The difference between the QG and SN errors is most pronounced at low measurement strength $\Omega_q$. It gradually decreases as the measurement strength increases — a consequence of the SN term $\propto \omega_{\rm SN}^2(\hat x - x_c)$ being suppressed by the measurement.  At certain regions of the measurement strength $\Omega_q$, the SN contribution to the tomography error can take negative values, which means the QG-filtered inference assigns a smaller quadrature variance than the corresponding initial-state contribution $(\bm{V}_0)_{ij}$.

Furthermore, in standard quantum mechanics, it is straightforward to show that $[\bm{V}^{\rm error}_{\rm QG}]_{xx} \sim \hbar\Omega_{\rm th}^{1/2}/M\Omega_q^{3/2}$, $[\bm{V}^{\rm error}_{\rm QG}]_{xp} \sim -\hbar\Omega_{\rm th}/\Omega_q$, $[\bm{V}^{\rm error}_{\rm QG}]_{pp} \sim \hbar M \Omega_{\rm th}^{3/2}/\Omega_q^{1/2}$, where $[\bm{V}^{\rm error}_{\rm QG}]_{xx}$ has the strongest dependence on $\Omega_q$.
As shown in Fig.~\ref{fig:error_components_temperature}, the contribution of the SN effect to the tomography error (denoted by dot-dashed lines) $\bm{V}^{\rm error}_{\rm SN} - \bm{V}^{\rm error}_{\rm QG}$ generally has a relatively smoother dependence on $\Omega_q$. Decreasing the measurement strength decreases the relative SN contribution $\bigl([\bm{V}^{\rm error}_{\rm SN} - \bm{V}^{\rm error}_{\rm QG}]_{xx}\bigr) /[\bm{V}^{\rm error}_{\rm QG}]_{xx}$ most rapidly, compared to the other two components.  In addition, comparing the two panels of Fig.~\ref{fig:error_components_temperature} shows that at higher temperatures, classical thermal noise reduces the SN-induced difference. This is because classical thermal noise does not affect the evolution of the conditional variance, and therefore does not affect $\sigma^2_{\rm SN}$ or $\bm{V}^{\rm error}_{\rm SN} - \bm{V}^{\rm error}_{\rm QG}$ but does increase $\bm{V}^{\rm error}_{\rm QG}$. Figure~\ref{fig:error_components} then takes a representative low-temperature slice across the region with negative ${\rm det}\bm V^{\rm recon}_{\rm SN}$ in Fig.~\ref{fig:SN_uncertainty_violation}. It illustrates that, around $\Omega_q\sim 0.1\,{\rm Hz}$, the $x$-$p$ cross-variance $[\bm{V}^{\rm error}_{\rm SN}]_{xp}$ is large and the momentum component $[\bm{V}^{\rm error}_{\rm SN}]_{pp}$ becomes negative, which means $\det(\bm{V}_{\rm SN}^{\rm recon})=[\bm{V}^{\rm error}_{\rm SN}]_{xx}[\bm{V}^{\rm error}_{\rm SN}]_{pp}-[\bm{V}^{\rm error}_{\rm SN}]^2_{xp}<0$.

 More generally, the QG and SN dynamics contribute differently to the reconstructed covariance error. This difference in the error covariance provides an operational way to distinguish the SN and QG hypotheses.

\section{Possible Experimental Signatures}\label{sec:experiment}
The SN-induced effects on the tomography process discussed in the previous section could leave possible signatures in an experiment. In this section, we focus on the most direct operational test: tomography of an initially prepared Gaussian test-mass state with the QG/standard-quantum reconstruction map held fixed. This choice keeps the data analysis common to the two hypotheses. Any difference between the QG and SN reconstructions then comes from the dynamics that generated the measurement record, rather than from using different estimators for the two theories. The signatures are threefold: the reconstructed SN covariance becomes angle-set dependent, the inferred covariance can leave the standard Gaussian-covariance domain under the mismatched QG map, and the QG--SN distance varies systematically with measurement strength and temperature.

Consider an optomechanical setup for performing tomography on the test mass quantum state $|\psi(0)\rangle$, which is a Gaussian pure state. Regardless of the underlying physics (QG or SN), we follow the state tomography approach of standard quantum mechanics, which means we reconstruct the covariance matrix of $|\psi(0)\rangle$ by using the optimal filter in QG to estimate the error at three mechanical quadrature angles\,$(0,\pi/4,\pi/2)$. According to the discussion in Sec.\,\ref{sec:reconstruction}, different underlying physics will lead to different reconstructed quantum states. Larger "distance" between the reconstructed states in the QG and SN cases means higher distinguishability of the two theories.

\begin{figure}[h]
\centering
\includegraphics[width=0.49\textwidth]{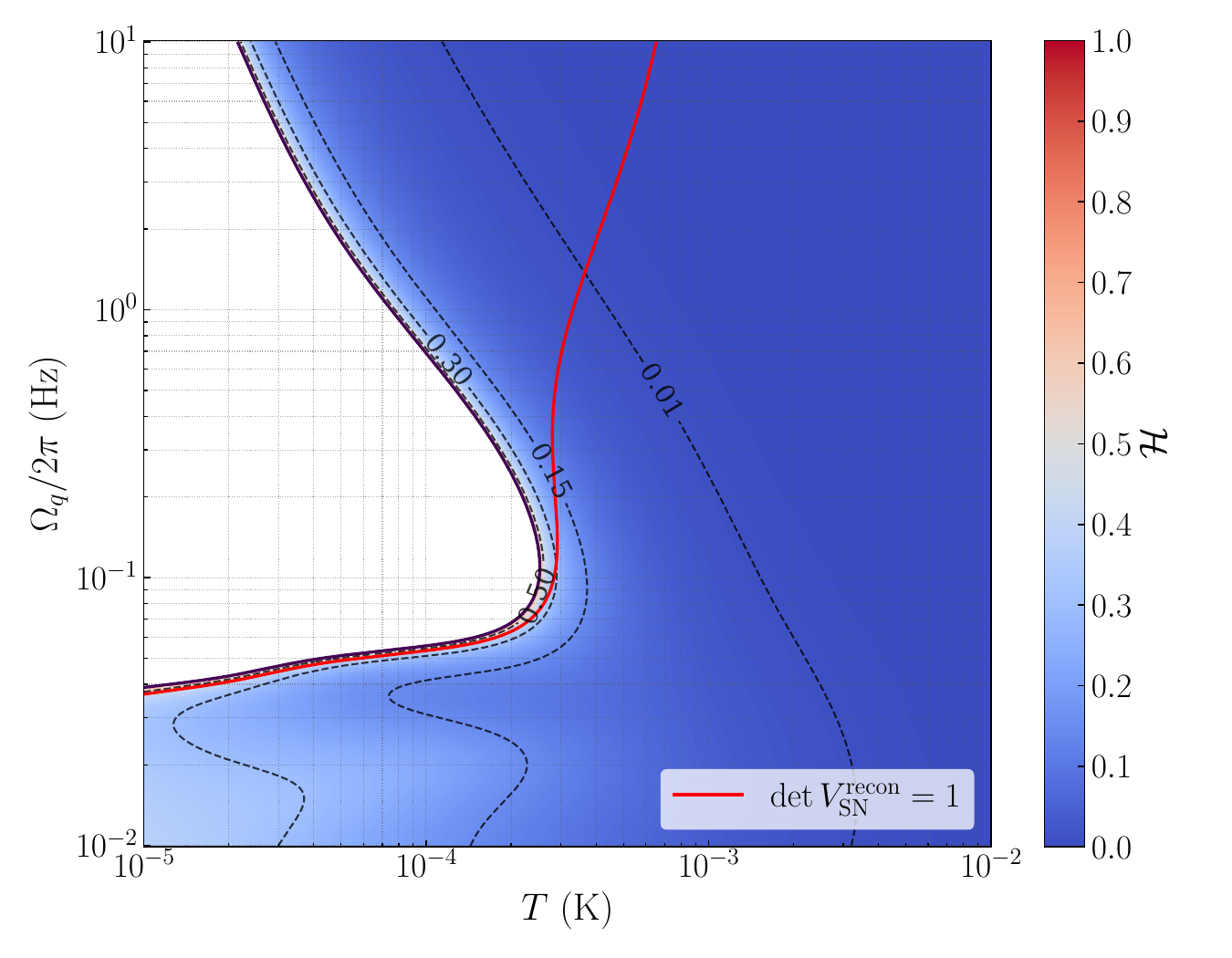}
\caption{Distinguishability of the QG and SN reconstructions in the $(T,\Omega_q)$ plane. The color scale gives the Hellinger distance $\mathcal{H}=\mathcal{H}(\bm{V}_{\rm QG}^{\rm recon},\bm{V}_{\rm SN}^{\rm recon})$, with dashed contours indicating selected values of $\mathcal{H}$. The red curve marks the determinant threshold for the reconstructed SN covariance in the dimensionless plotting convention. White regions correspond to parameter points for which the QG-filtered SN reconstruction lies outside the standard Gaussian-covariance domain.}
\label{fig:fidelity_contour}
\end{figure}

The distance between two reconstructed Gaussian covariance matrices $(\bm{\Sigma}_1,\bm{\Sigma}_2)$ can be formulated as the so-called Hellinger distance:
\begin{equation}\label{eq:geometric_distinguishability}
  \mathcal H (\bm{\Sigma}_1,\bm{\Sigma}_2)
  =
  1-
  \frac{
    \det(\bm{\Sigma}_1)^{1/4}\det(\bm{\Sigma}_2)^{1/4}
  }{
    \det\!\left[(\bm{\Sigma}_1+\bm{\Sigma}_2)/2\right]^{1/2}
  } .
\end{equation}
Taking
$\bm{\Sigma}_1=\bm{V}_{\rm QG}^{\rm recon}$ and
$\bm{\Sigma}_2=\bm{V}_{\rm SN}^{\rm recon}$, the Hellinger distance across the parameter space is shown in Fig.~\ref{fig:fidelity_contour}. The regions where the QG-filtered SN covariance lies outside the standard Gaussian-covariance domain are left blank. 

\begin{figure*}
\centering
\includegraphics[width=0.8\textwidth]{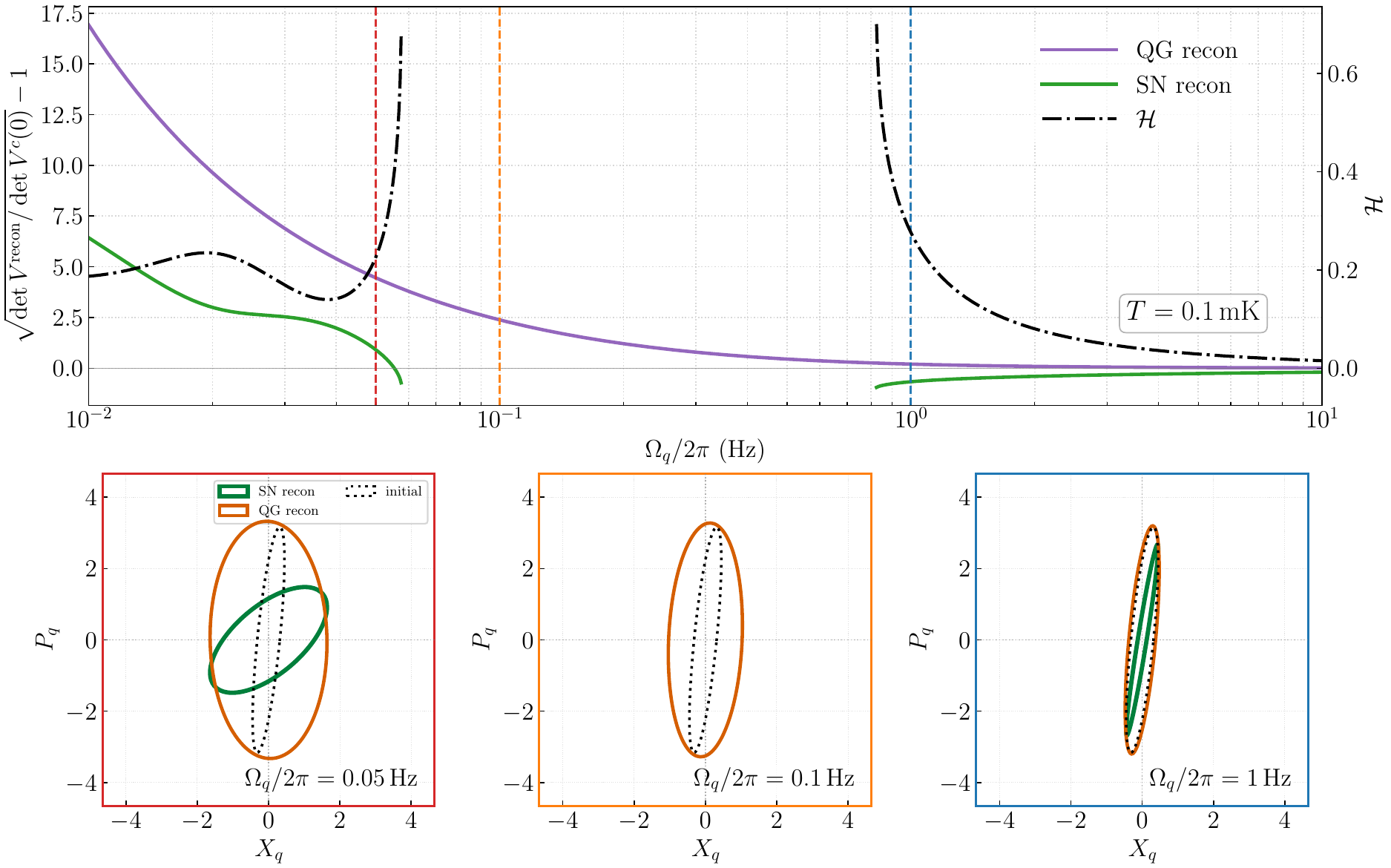}
\caption{Low-temperature slice of the reconstruction at $T=0.1$ mK. The top panel shows the deviation of the $\sqrt{\det\bm{V}^{\rm recon}/\det\bm{V}^c(0)}-1$ for QG and SN reconstructed state from that of the initial state, together with the Hellinger distance $\mathcal{H}$. The lower panels show reconstructed phase-space uncertainty ellipses at three representative measurement strengths, indicated by the vertical dashed lines above: SN in green, QG in orange, and the initial covariance in black dotted lines. In this low-noise regime, the SN correction can reduce the inferred covariance and push $\bm{V}_{\rm SN}^{\rm recon}$ outside the standard Gaussian-covariance domain over part of the scan.}
\label{fig:noise_det_lowT}
\end{figure*}

\begin{figure*}
\centering
\includegraphics[width=0.8\textwidth]{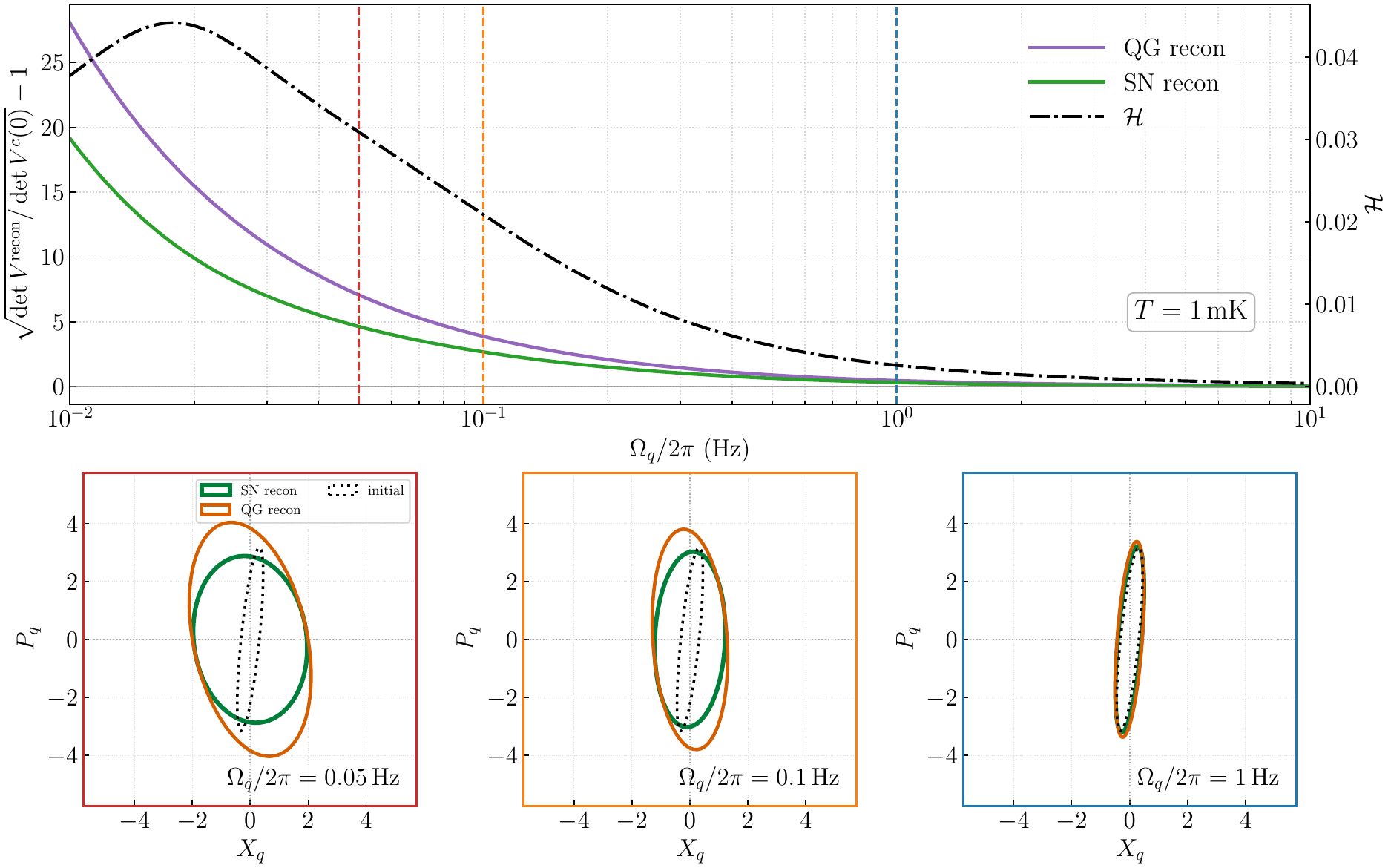}
\caption{High-temperature slice of the reconstruction at $T=1$ mK. The plotted quantities and line styles are the same as in Fig.~\ref{fig:noise_det_lowT}. Thermal noise now dominates the reconstruction error, so the QG and SN uncertainty ellipses remain closer to one another and the Hellinger distance is smaller. The enhanced thermal broadening also removes the below-boundary determinant region that appears in the low-temperature case.}
\label{fig:noise_det_highT}
\end{figure*}

We also take $T=0.1$\,mK and $T=1$\,mK as the low- and high-temperature slices of Fig.~\ref{fig:fidelity_contour}, and plot the uncertainty ellipses of the reconstructed state at selected measurement strengths in Fig.\,\ref{fig:noise_det_lowT} and Fig.\,\ref{fig:noise_det_highT}. In doing so, we use the dimensionless coordinates $(X_q, P_q )\equiv(x/\sqrt{\hbar/(2M\omega_q)}, p/\sqrt{\hbar M\omega_q/2})$.

For the low-temperature case $T = 0.1$ mK, the SN effect is stronger than in the relatively higher-temperature case $T = 1$ mK. In the low-temperature scenario, there exists a parameter region where $\det[\bm{V}_{\rm SN}^{\rm recon}] < 0$, as discussed in the previous section. In this region, the QG-filtered SN covariance is outside the domain on which the Gaussian Hellinger distance in Eq.~\eqref{eq:geometric_distinguishability} is defined, so the corresponding points are excluded from the contour plot. As the measurement strength increases, the reconstructed state under SN dynamics becomes increasingly similar to that under QG dynamics, thereby leading to a decrease in $\mathcal{H}$, because a strong measurement suppresses the SN effect. In contrast, while a weak measurement leads to noisy tomography in the QG case, the SN correction can reduce the inferred mechanical quadrature uncertainties, resulting in a non-negligible Hellinger distance wherever the reconstructed covariance remains inside the standard Gaussian-covariance domain. For the high-temperature case, there is no excluded region where $\det\bm{V}^{\rm recon}_{\rm SN}<0$ due to the enhanced thermal noise; the SN correction is instead masked by thermal broadening, so the distinguishability changes smoothly across the scanned parameter range.

These results identify the useful parameter regime for the fixed-map test. The signal is not enhanced simply by increasing the measurement strength. A very strong measurement reduces the conditional covariance and thereby suppresses the SN term itself, while high thermal noise hides the model mismatch under an ordinary positive tomographic error. The most informative region is therefore an intermediate one: the measurement must be strong enough to perform tomography, but not so strong that the conditional state is forced into the regime where the SN correction becomes negligible.

The experimental requirements identified here remain extremely demanding. The key step before experimentally probing the SN influence on tomography process is to first preparing macroscopic test mass quantum state in the mHz range, which is already very challenging due to the stringent requirement on a sufficiently high-$Q$ oscillator operating at mHz-scale mechanical frequencies, a regime where maintaining low mechanical loss is difficult, together with millikelvin-scale temperature. This suggests that continued progress in low-frequency high-Q test mass and cryogenic operation will be central for approaching the regime considered here; optical readout loss would also need to be kept sufficiently low in an experimental implementation. In this route, recent progress in low-frequency optomechanical sensing provides promising experimental platforms such as TOBAR\,\cite{Ando2010}.

\section{Tomography in General Nonlinear Quantum Mechanics}\label{sec:nonlinear_tomography}
The SN example above suggests a more general lesson for tomography in nonlinear quantum mechanics\,\cite{BialynickiBirula1976,Weinberg1989,Gisin1990,Polchinski1991,Bassi_2017}. In this section, we schematically formulate that lesson as a general finite-time tomography structure for state-dependent dynamics, and then identify the SN case as one concrete realization. By nonlinear quantum mechanics (NLQM) we mean a theory in which the physical dynamical law is not a linear map on the quantum state. For a pure state, this can be represented schematically as
\begin{equation}\label{eq:general_pure_NLQM}
  i\hbar\frac{d}{dt}\ket{\psi_t}
  =
  \left[
  \hat H_0+\hat H_{\rm NL}[\psi_t,\psi_t^*]
  \right]\ket{\psi_t},
\end{equation}
where $\hat H_{\rm NL}$ is a state-dependent Hamiltonian functional. Equivalently, for a density matrix, one may write
\begin{equation}\label{eq:general_density_NLQM}
  \dot\rho_t
  =
  \mathcal L_0[\rho_t]
  +\mathcal N_{\rm NL}[\rho_t],
\end{equation}
where $\mathcal L_0$ is the ordinary linear quantum generator and $\mathcal N_{\rm NL}$ is a nonlinear dynamical functional satisfying, in general,
\begin{equation}
  \mathcal N_{\rm NL}[p\rho_1+(1-p)\rho_2]
  \neq
  p\mathcal N_{\rm NL}[\rho_1]
  +(1-p)\mathcal N_{\rm NL}[\rho_2].
\end{equation}
The relevant point for tomography is not merely that a normalized conditional state obeys a nonlinear stochastic filtering equation; this already happens in standard quantum mechanics. The important distinction is whether the physical nonlinear term $\mathcal N_{\rm NL}$ acts during the readout and depends on the state parameters that tomography is trying to infer. If it does, then the measurement is no longer a passive projection of a pre-existing state through a state-independent measurement kernel.

The results in Secs.~\ref{sec:reconstruction} and~\ref{sec:experiment} are a concrete realization of this general structure. In the QG/standard-quantum case, the fixed reconstruction map has the ordinary linear form: different angle sets reconstruct the same covariance matrix, the added error is a positive noise covariance, and the Hellinger distance from an SN-generated reconstruction vanishes when the nonlinear correction is suppressed. In the SN case, the nonlinear correction is generated by the state-dependent self-gravity term and by the conditional covariance evolution. Thus the three signatures found above---angle-set dependence, below-boundary QG-filtered covariances, and a measurement-strength/temperature dependent distinguishability---are not isolated numerical artifacts. They realize the following more general NLQM statement: when the physical dynamics acting during readout depend on the state parameters being inferred, finite-time tomography can inherit a nonlinear, state-dependent correction to its reconstruction map.

Let $\bm v_0$ denote the set of initial state parameters to be reconstructed. For Gaussian covariance tomography, for example,
\begin{equation}
  \bm v_0=(V_{xx}(0),V_{xp}(0),V_{pp}(0))^T .
\end{equation}
In ordinary linear Gaussian tomography, the filtered variance for a setting $a$ can be written in the affine form
\begin{equation}\label{eq:linear_general_tomography}
  S_a=\bm r_a^T\bm v_0+N_a ,
\end{equation}
where $\bm r_a$ is the response vector and $N_a$ is the residual estimation error for the chosen filter. In the protocol considered in this paper, this residual is not subtracted after the measurement. Instead, one chooses the filter so that the residual contribution is minimized. For a set of three quadrature angles $\Theta=\{a_1,a_2,a_3\}$, Eq.~\eqref{eq:linear_general_tomography} gives the operational reconstruction
\begin{equation}\label{eq:linear_general_reconstruction}
  \bm v_{\rm rec}^{(\Theta)}
  =
  R_\Theta^{-1}\bm S_\Theta
  =
  \bm v_0+R_\Theta^{-1}\bm N_\Theta,
  \qquad
  R_\Theta=
  \begin{pmatrix}
  \bm r_{a_1}^T\\
  \bm r_{a_2}^T\\
  \bm r_{a_3}^T
  \end{pmatrix}.
\end{equation}
The first term is the covariance vector being reconstructed, while $R_\Theta^{-1}\bm N_\Theta$ is the covariance contribution from the residual estimation error. When the reconstruction model is correct, different nonsingular choices of $\Theta$ should therefore be compatible with one covariance matrix, up to this optimized residual contribution. This is the finite-dimensional version of the usual Radon consistency of optical tomography.

In a nonlinear quantum theory, the state during the measurement is generated by a state-dependent law. Schematically, one may write the conditioned evolution and measurement record as
\begin{align}
  d\rho_t
  &=
  \left[
  \mathcal L_{u(t)}[\rho_t]
  +\mathcal N_{{\rm NL},u(t)}[\rho_t]
  \right]dt
  +\mathcal M_{u(t)}[\rho_t]\,dW_t, \label{eq:general_nonlinear_filter}\\
  dY_t
  &=
  m_{u(t)}[\rho_t]\,dt+dW_t ,
\end{align}
where $u(t)$ denotes the chosen measurement setting or control. The term $\mathcal L_u$ is the linear quantum-mechanical part of the open-system dynamics, $\mathcal M_u$ is the stochastic measurement update, and $\mathcal N_{{\rm NL},u}$ is the nonlinear effect. The subscript $u$ allows for the fact that the apparent nonlinear correction during readout can depend on the measurement setting, for example through the conditional trajectory selected by the measurement. After filtering the record, the measured variance generally has the structure
\begin{equation}\label{eq:nonlinear_general_tomography}
  S_a
  =
  \bm r_a^T\bm v_0
  +N_a
  +\Delta_a[\bm v_0;u,\mathcal N_{\rm NL}] .
\end{equation}
Here $\Delta_a$ denotes the finite-time tomographic correction induced by the nonlinear dynamical functional $\mathcal N_{\rm NL}$. This last term is the essential difference from Eq.~\eqref{eq:linear_general_tomography}. It is a nonlinear functional of the same unknown parameters that the experiment is attempting to reconstruct, and it also depends on the measurement setting used during the readout.

If one nevertheless applies the ordinary linear reconstruction map to data generated by Eq.~\eqref{eq:nonlinear_general_tomography}, the result is
\begin{equation}\label{eq:nonlinear_general_reconstruction}
  \bm v_{\rm rec,NL}^{(\Theta)}
  =
  \bm v_0
  +
  R_\Theta^{-1}\bm N_\Theta
  +
  R_\Theta^{-1}\bm\Delta_\Theta[\bm v_0;u,\mathcal N_{\rm NL}],
\end{equation}
where $\bm\Delta_\Theta=(\Delta_{a_1},\Delta_{a_2},\Delta_{a_3})^T$. Thus the residual estimation error contribution remains present as $R_\Theta^{-1}\bm N_\Theta$, while the genuinely nonlinear contribution is the additional term $R_\Theta^{-1}\bm\Delta_\Theta$. Equation~\eqref{eq:nonlinear_general_reconstruction} makes the main point explicit. In linear quantum mechanics $\bm\Delta_\Theta=0$, so the residual error is the optimized finite-time measurement contribution. In a nonlinear theory, the correction is both state-dependent and setting-dependent, so two different angle sets generally give
\begin{equation}
  \bm v_{\rm rec,NL}^{(\Theta)}
  -
  \bm v_{\rm rec,NL}^{(\Theta')}
  =  R_\Theta^{-1}\bm\Delta_\Theta
  -
  R_{\Theta'}^{-1}\bm\Delta_{\Theta'} ,
\end{equation}
which need not vanish. In the QG construction used in the previous sections, the optimized residual term has the usual covariance interpretation and is minimized by properly choosing the filter. The new NLQM feature is that the additional $\bm\Delta_\Theta$ term is not an independently calibratable residual error: it depends on the state parameters being inferred and on the measurement setting itself. This is the formal origin of angle-set-dependent reconstructed covariances.

The same statement can be phrased in terms of the Radon transform. In ordinary quadrature tomography, the measured marginal at angle $\varphi$ is
\begin{equation}
  p_\varphi(X)
  =
  \int dx\,dp\,W_0(x,p)\,
  \delta\!\left(X-x\cos\varphi-p\sin\varphi\right),
\end{equation}
or, equivalently, its characteristic function satisfies the central-slice relation
\begin{equation}
  \chi_\varphi(k)=\chi_W(k\cos\varphi,k\sin\varphi).
\end{equation}
Here
\begin{align}
  \chi_\varphi(k)
  &=
  \int dX\,e^{ikX}p_\varphi(X),
  \\
  \chi_W(\xi_x,\xi_p)
  &=
  \int dx\,dp\,e^{i(\xi_x x+\xi_p p)}W_0(x,p)
\end{align}
are, respectively, the one-dimensional characteristic function of the measured quadrature marginal and the phase-space characteristic function of the Wigner distribution.
We introduce $\chi_\varphi$ because the central-slice theorem is most transparent in characteristic-function form: it expresses tomography as the statement that all measured marginals must come from one underlying phase-space distribution. In the nonlinear case, this is exactly the consistency condition that can fail.
For a finite-time measurement in a nonlinear theory, the experimentally obtained marginal is instead of the form
\begin{equation}
  \tilde p_\varphi(X)
  =
  \mathcal T_\varphi^{\rm NL}[W_0](X),
\end{equation}
where $\mathcal T_\varphi^{\rm NL}$ includes both the measurement back action and the state-dependent dynamics during the readout. Unless there exists a single angle-independent phase-space distribution $W_{\rm eff}$ whose Radon projections reproduce all the $\tilde p_\varphi$, the inverse Radon transform is only a formal inversion and not a reconstructed Wigner function of one underlying state.

For Gaussian covariance tomography, this Radon-consistency condition has an especially simple form. A true covariance matrix implies
\begin{equation}
  V_\varphi
  =
  V_{xx}\cos^2\varphi
  +V_{pp}\sin^2\varphi
  +2V_{xp}\sin\varphi\cos\varphi ,
\end{equation}
so the angle dependence must be expressible as
\begin{equation}
  V_\varphi=A+B\cos2\varphi+C\sin2\varphi .
\end{equation}
The nonlinear correction $\Delta_\varphi[\bm v_0;u,\mathcal N_{\rm NL}]$ in Eq.~\eqref{eq:nonlinear_general_tomography} does not have this form. It may depend on the full measurement history, on the chosen filter, and on the conditional trajectory generated by the nonlinear dynamics. Then separately optimized measurements at different angles can produce Gaussian-looking marginals that are not projections of any single covariance matrix.

The SN theory is a particularly important and concrete example of this general mechanism. It is not introduced here as an arbitrary nonlinear modification, but as the non-relativistic limit of semi-classical gravity with a substantial literature on its foundations, criticisms, regularizations, and phenomenology\,\cite{Bahrami_2014,Anastopoulos_2014,Diosi1989,Penrose1996,Giulini2011,Giulini2014,Grossardt2016,Gan2016Optomechanical}. At the same time, for the Gaussian optomechanical setting studied in this paper it is tractable enough to compute the nonlinear correction explicitly. In the notation of Sec.~\ref{sec:reconstruction}, the abstract nonlinear generator $\mathcal N_{\rm NL}$ is the SN state-dependent self-gravity drift, and the abstract tomographic correction $\Delta_a$ is realized by
\begin{equation}\label{eq:SN_as_general_delta}
  \Delta_\theta^{\rm SN}
  =
  \sigma_{\rm SN}^2[g_1,g_2;h_i(t;h_j(0))],
\end{equation}
where the functions $h_i(t)$ are determined by the Riccati evolution of the conditional covariance. Thus the SN theory provides a worked example in which the nonlinear tomographic correction can be calculated, plotted, and compared directly with the QG/standard-quantum reconstruction.

This general argument also clarifies the limits of the claim. A nonlinear theory does not automatically produce a large tomographic signal. If the nonlinear term is negligible during the readout, if it reduces over the relevant regime to a known state-independent parameter shift, or if the measurement is effectively instantaneous compared with the nonlinear evolution, then $\Delta_a$ may be absent or experimentally indistinguishable from a calibration change. The formal statement established here is therefore limited but clear: when appreciable state-dependent nonlinear dynamics acts during a finite-time tomographic readout, the standard linear reconstruction map can become model dependent and can fail the usual angle-set or Radon-consistency tests. The SN analysis in this paper is one explicit worked example of that structure, rather than a proof that all nonlinear quantum theories display the same signatures with the same strength.

\section{Discussion and Conclusion}\label{sec:conclusion}
We have studied continuous optomechanical tomography of a macroscopic test mass oscillator in the presence of Schr{\"o}dinger-Newton self-gravity. In the QG reference model, the optimal filter gives a positive tomographic error covariance, and the reconstruction of a physical Gaussian initial state remains inside the standard Gaussian-covariance domain. In the SN theory, the measurement record is generated by a conditional covariance dynamics that depends on the state-dependent self-gravity term. Applying the same fixed QG reconstruction map to such a record produces an additional contribution that depends on the homodyne angle and on the conditional covariance evolution.

This leads to three main features of the fixed-map comparison. First, the reconstruction in the SN case depends on the chosen set of tomographic quadrature angles, unlike the QG reconstruction. Second, the QG-filtered SN covariance can leave the standard Gaussian-covariance domain in some regions of measurement strength and temperature. This is a diagnostic that the assumed QG reconstruction model is incompatible with the dynamics that generated the record, and it could be used as an operational feature to distinguish QG and SN theories in a state-verification experiment. Third, the Hellinger distance between QG and SN reconstructions exhibits nontrivial dependence on temperature and measurement strength. Low temperature and moderate measurement strength enhance the model mismatch, while strong measurement strength suppresses the SN contribution and high thermal noise masks it.

The general nonlinear-tomography discussion in Sec.~\ref{sec:nonlinear_tomography} also clarifies why a fully self-consistent SN tomography protocol for an unknown initial state is not obtained by the simple replacement $\sigma_{\rm QG}^2\to\sigma_{\rm QG}^2+\sigma_{\rm SN}^2$ in the standard filter optimization. In QG tomography the filtered variance has the schematic form
\begin{equation}
  \sigma_Y^2(\theta)
  =
  \alpha^2\bm{\lambda}^T(\theta)\bm v_0
  +\sigma_{\rm QG}^2(\theta),
\end{equation}
where $\bm v_0=(V_{xx}(0),V_{xp}(0),V_{pp}(0))^T$ is the unknown initial covariance and $\sigma_{\rm QG}^2$ is a calculable noise term independent of $\bm v_0$. In SN theory the corresponding expression is instead
\begin{equation}
  \sigma_Y^2(\theta)
  =
  \alpha^2\bm{\lambda}^T(\theta)\bm v_0
  +\sigma_{\rm QG}^2(\theta)
  +\mathcal F_\theta[h_i(t;h_j(0))],
\end{equation}
with $\mathcal F_\theta=\sigma_{\rm SN}^2$ in the notation of Sec.~\ref{sec:reconstruction}. Since the initial values $h_j(0)$ are just another parametrization of the unknown covariance, the SN correction contains the same unknown object as the signal term. Thus the problem becomes a nonlinear inverse problem for the whole measurement record, not three independent linear equations for three covariance components.

The same point appears in the Radon transformation. If $h_i(0)$ were independently known, one could propagate $h_i(t)$ and design a filter for that specified state-preparation hypothesis. For tomography of an unknown state, however, different homodyne angles and filters generate different conditional covariance trajectories. The resulting set of measured quadrature distributions need not be the Radon projections of a single Wigner function, even if each marginal looks Gaussian. This is the SN realization of Eq.~\eqref{eq:nonlinear_general_reconstruction}: the correction term depends on both the state and the measurement setting.

These observations do not mean that self-consistent SN tomography is impossible. They mean that it must be formulated as a nonlinear statistical estimation problem. A possible future route is Bayesian or adaptive estimation: assign candidate initial covariances, propagate the SN conditional equations for each candidate under the chosen measurement schedule, and update the candidate weights from the observed record. Whether this procedure is identifiable, stable, and experimentally efficient remains open, and we leave it for future work. The fixed-QG-map signatures studied here should therefore be read as the main operational result of the present paper, while optimal tomography internal to SN theory is a separate problem.

The broader message is that SN theory is both physically motivated and technically instructive as a specific example of nonlinear quantum mechanics. In the finite-time fixed-map setting studied here, it shows explicitly how state-dependent dynamics during readout can modify the tomographic map itself, leading to angle-set dependence, below-boundary QG-filtered covariances under the standard Gaussian interpretation, and Radon-consistency obstructions. Whether these features appear in every nonlinear modification of quantum mechanics remains an open question, but the formal structure developed in Sec.~\ref{sec:nonlinear_tomography} shows why they are natural diagnostics whenever appreciable state-dependent nonlinear dynamics is present during a finite-duration tomography experiment.

\acknowledgements
Y.\,M. W.\,Z. and Y.\,L. is supported by the National Natural Science Foundation of China under Grant No.12474481 and No. 12441503, National Key R$\&$D Program of China (2023YFC2205801). Y. C. is supported by Simons Foundation and the Keck Foundation. H. M. is supported by National Natural Science Foundation of China under Grant No. 12441503 and National Key R$\&$D Program of China (2023YFC2205800).

\bibliographystyle{unsrt}
\bibliography{causal-conditional}

@article{Ando2010,
    author = "Ando, Masaki and Ishidoshiro, Koji and Yamamoto, Kazuhiro and Yagi, Kent and Kokuyama, Wataru and Tsubono, Kimio and Takamori, Akiteru",
    title = "{Torsion-Bar Antenna for Low-Frequency Gravitational-Wave Observations}",
    doi = "10.1103/PhysRevLett.105.161101",
    journal = "Phys. Rev. Lett.",
    volume = "105",
    pages = "161101",
    year = "2010"
}

@ARTICLE{Vyatchanin1996,
       author = {{Vyatchanin}, S.~P. and {Matsko}, A.~B.},
        title = "{Quantum variational force measurement and the cancellation of nonlinear feedback}",
      journal = {Soviet Journal of Experimental and Theoretical Physics},
         year = 1996,
        month = jun,
       volume = {82},
       number = {6},
        pages = {1007-1014},
       adsurl = {https://ui.adsabs.harvard.edu/abs/1996JETP...82.1007V}
}

@article{Regal2017,
  title = {Improving Broadband Displacement Detection with Quantum Correlations},
  author = {Kampel, N. S. and Peterson, R. W. and Fischer, R. and Yu, P.-L. and Cicak, K. and Simmonds, R. W. and Lehnert, K. W. and Regal, C. A.},
  journal = {Phys. Rev. X},
  volume = {7},
  issue = {2},
  pages = {021008},
  numpages = {10},
  year = {2017},
  month = {Apr},
  publisher = {American Physical Society},
  doi = {10.1103/PhysRevX.7.021008},
  url = {https://link.aps.org/doi/10.1103/PhysRevX.7.021008}
}

@ARTICLE{Feng2022,
       author = {{Feng}, Tianfeng and {Vedral}, Vlatko},
        title = "{Amplification of gravitationally induced entanglement}",
      journal = {Physical Review D},
         year = 2022,
        month = sep,
       volume = {106},
       number = {6},
          eid = {066013},
        pages = {066013},
          doi = {10.1103/PhysRevD.106.066013},
archivePrefix = {arXiv},
       eprint = {2202.09737},
 primaryClass = {quant-ph},
       adsurl = {https://ui.adsabs.harvard.edu/abs/2022PhRvD.106f6013F}
}

@ARTICLE{Podzien2026,
       author = {{P{\l}odzie{\'n}}, Marcin and {Os{\k{e}}ka-Lenart}, Julia and {Lewenstein}, Maciej and {Eckstein}, Micha{\l}},
        title = "{Entanglement generation in a two-body Schr{\"o}dinger--Newton model}",
      journal = {arXiv e-prints},
         year = 2026,
        month = may,
          eid = {arXiv:2605.06577},
        pages = {arXiv:2605.06577},
          doi = {10.48550/arXiv.2605.06577},
archivePrefix = {arXiv},
       eprint = {2605.06577},
 primaryClass = {quant-ph},
       adsurl = {https://ui.adsabs.harvard.edu/abs/2026arXiv260506577P}
}

@ARTICLE{Kimble2001,
       author = {{Kimble}, H.~J. and {Levin}, Yuri and {Matsko}, Andrey B. and {Thorne}, Kip S. and {Vyatchanin}, Sergey P.},
        title = "{Conversion of conventional gravitational-wave interferometers into quantum nondemolition interferometers by modifying their input and/or output optics}",
      journal = {\prd},
         year = 2001,
        month = dec,
       volume = {65},
       number = {2},
          eid = {022002},
        pages = {022002},
          doi = {10.1103/PhysRevD.65.022002},
archivePrefix = {arXiv},
       eprint = {gr-qc/0008026},
 primaryClass = {gr-qc},
       adsurl = {https://ui.adsabs.harvard.edu/abs/2001PhRvD..65b2002K}
}

@article{Wieczork2015,
  title = {Optimal State Estimation for Cavity Optomechanical Systems},
  author = {Wieczorek, Witlef and Hofer, Sebastian G. and Hoelscher-Obermaier, Jason and Riedinger, Ralf and Hammerer, Klemens and Aspelmeyer, Markus},
  journal = {Phys. Rev. Lett.},
  volume = {114},
  issue = {22},
  pages = {223601},
  numpages = {6},
  year = {2015},
  month = {Jun},
  publisher = {American Physical Society},
  doi = {10.1103/PhysRevLett.114.223601},
  url = {https://link.aps.org/doi/10.1103/PhysRevLett.114.223601}
}

@article{Liu2026,
  title = {Testing the quantum nature of gravity through interferometry},
  author = {Liu, Yubao and Chen, Yanbei and Somiya, Kentaro and Ma, Yiqiu},
  journal = {Phys. Rev. D},
  volume = {113},
  issue = {2},
  pages = {022002},
  numpages = {36},
  year = {2026},
  month = {Jan},
  publisher = {American Physical Society},
  doi = {10.1103/15n3-zzjw},
  url = {https://link.aps.org/doi/10.1103/15n3-zzjw}
}

@ARTICLE{Miao2010,
       author = {{Miao}, Haixing and {Danilishin}, Stefan and {M{\"u}ller-Ebhardt}, Helge and {Rehbein}, Henning and {Somiya}, Kentaro and {Chen}, Yanbei},
        title = "{Probing macroscopic quantum states with a sub-Heisenberg accuracy}",
      journal = {\pra},
         year = 2010,
        month = jan,
       volume = {81},
       number = {1},
          eid = {012114},
        pages = {012114},
          doi = {10.1103/PhysRevA.81.012114},
archivePrefix = {arXiv},
       eprint = {0905.3729},
 primaryClass = {quant-ph},
       adsurl = {https://ui.adsabs.harvard.edu/abs/2010PhRvA..81a2114M}
}

@article{Groshardt2022,
       author = {{Gro{\ss}ardt}, Andr{\'e}},
        title = "{Three little paradoxes: Making sense of semiclassical gravity}",
      journal = {AVS Quantum Science},
         year = 2022,
        month = mar,
       volume = {4},
       number = {1},
          eid = {010502},
        pages = {010502},
          doi = {10.1116/5.0073509},
archivePrefix = {arXiv},
       eprint = {2201.10452},
 primaryClass = {gr-qc},
       adsurl = {https://ui.adsabs.harvard.edu/abs/2022AVSQS...4a0502G}
}

@BOOK{Mueller1962,
       author = {{Mueller C}},
        title = "{Les The'ories Relativistes de la Gravitation (Colloques Internationaux CNRS), edited by A Lichnerowicz and M-A Tonnelat}",
         year = 1962,
       publisher = {Paris: CNRS}
}

@article{Rosenfeld1963,
	Adsurl = {https://ui.adsabs.harvard.edu/abs/1963NucPh..40..353R},
	Author = {{Rosenfeld}, L.},
	Doi = {10.1016/0029-5582(63)90279-7},
	Journal = {Nuclear Physics},
	Month = feb,
	Pages = {353-356},
	Title = {{On quantization of fields}},
	Volume = {40},
	Year = 1963}

@article{Bahrami_2014,
  doi = {10.1088/1367-2630/16/11/115007},
  url = {https://doi.org/10.1088/1367-2630/16/11/115007},
  year = 2014,
  month = {nov},
  publisher = {{IOP} Publishing},
  volume = {16},
  number = {11},
  pages = {115007},
  author = {Mohammad Bahrami and Andr{\'{e}} Gro{\ss}ardt and Sandro Donadi and Angelo Bassi},
  title = {The Schrödinger{\textendash}Newton equation and its foundations},
  journal = {New Journal of Physics},
}

@article{Bose2017,
  title = {Spin Entanglement Witness for Quantum Gravity},
  author = {Bose, Sougato and Mazumdar, Anupam and Morley, Gavin W. and Ulbricht, Hendrik and Toro\ifmmode \check{s}\else \v{s}\fi{}, Marko and Paternostro, Mauro and Geraci, Andrew A. and Barker, Peter F. and Kim, M. S. and Milburn, Gerard},
  journal = {Phys. Rev. Lett.},
  volume = {119},
  issue = {24},
  pages = {240401},
  numpages = {6},
  year = {2017},
  month = {Dec},
  publisher = {American Physical Society},
  doi = {10.1103/PhysRevLett.119.240401},
  url = {https://link.aps.org/doi/10.1103/PhysRevLett.119.240401}
}

@article{Marletto2017Gravitationally,
  title = {Gravitationally Induced Entanglement between Two Massive Particles is Sufficient Evidence of Quantum Effects in Gravity},
  author = {Marletto, C. and Vedral, V.},
  journal = {Phys. Rev. Lett.},
  volume = {119},
  issue = {24},
  pages = {240402},
  numpages = {5},
  year = {2017},
  month = {Dec},
  publisher = {American Physical Society},
  doi = {10.1103/PhysRevLett.119.240402},
  url = {https://link.aps.org/doi/10.1103/PhysRevLett.119.240402}
}

@article{Carney2021Using,
  title = {Using an Atom Interferometer to Infer Gravitational Entanglement Generation},
  author = {Carney, Daniel and M\"uller, Holger and Taylor, Jacob M.},
  journal = {PRX Quantum},
  volume = {2},
  issue = {3},
  pages = {030330},
  numpages = {16},
  year = {2021},
  month = {Aug},
  publisher = {American Physical Society},
  doi = {10.1103/PRXQuantum.2.030330},
  url = {https://link.aps.org/doi/10.1103/PRXQuantum.2.030330}
}

@article{Christodoulou2023,
  title = {Locally Mediated Entanglement in Linearized Quantum Gravity},
  author = {Christodoulou, Marios and Di Biagio, Andrea and Aspelmeyer, Markus and Brukner, \ifmmode \check{C}\else \v{C}\fi{}aslav and Rovelli, Carlo and Howl, Richard},
  journal = {Phys. Rev. Lett.},
  volume = {130},
  issue = {10},
  pages = {100202},
  numpages = {7},
  year = {2023},
  month = {Mar},
  publisher = {American Physical Society},
  doi = {10.1103/PhysRevLett.130.100202},
  url = {https://link.aps.org/doi/10.1103/PhysRevLett.130.100202}
}

@article{krisnanda2020,
  title={Observable quantum entanglement due to gravity},
  author={Krisnanda, Tanjung and Tham, Guo Yao and Paternostro, Mauro and Paterek, Tomasz},
  journal={npj Quantum Information},
  volume={6},
  number={1},
  pages={12},
  year={2020},
  publisher={Nature Publishing Group UK London},
  doi={10.1038/s41534-020-0243-y},
  url={https://www.nature.com/articles/s41534-020-0243-y}
}

@article{Wald2020,
	Adsurl = {https://ui.adsabs.harvard.edu/abs/2020IJMPD..2941003W},
	Author = {{Wald}, Robert M.},
	Doi = {10.1142/S0218271820410035},
	Eid = {2041003},
	Journal = {International Journal of Modern Physics D},
	Month = jan,
	Number = {11},
	Pages = {2041003},
	Title = {{Quantum superposition of massive bodies}},
	Volume = {29},
	Year = 2020}

@article{Belenchia2018,
	Adsurl = {https://ui.adsabs.harvard.edu/abs/2018PhRvD..98l6009B},
	Archiveprefix = {arXiv},
	Author = {{Belenchia}, Alessio and {Wald}, Robert M. and {Giacomini}, Flaminia and {Castro-Ruiz}, Esteban and {Brukner}, {\v{C}}aslav and {Aspelmeyer}, Markus},
	Doi = {10.1103/PhysRevD.98.126009},
	Eid = {126009},
	Eprint = {1807.07015},
	Journal = {\prd},
	Month = dec,
	Number = {12},
	Pages = {126009},
	Primaryclass = {quant-ph},
	Title = {{Quantum superposition of massive objects and the quantization of gravity}},
	Volume = {98},
	Year = 2018}

@article{Page1981,
	Author = {Page, Don N. and Geilker, C. D.},
	Doi = {10.1103/PhysRevLett.47.979},
	Issue = {14},
	Journal = {Phys. Rev. Lett.},
	Month = {Oct},
	Numpages = {0},
	Pages = {979--982},
	Publisher = {American Physical Society},
	Title = {Indirect Evidence for Quantum Gravity},
	Url = {https://link.aps.org/doi/10.1103/PhysRevLett.47.979},
	Volume = {47},
	Year = {1981}}

@article{Anastopoulos_2014,
	Author = {C Anastopoulos and B L Hu},
	Doi = {10.1088/1367-2630/16/8/085007},
	Journal = {New Journal of Physics},
	Month = {aug},
	Number = {8},
	Pages = {085007},
	Publisher = {{IOP} Publishing},
	Title = {Problems with the Newton{\textendash}Schr{\"o}dinger equations},
	Url = {https://doi.org/10.1088/1367-2630/16/8/085007},
	Volume = {16},
	Year = 2014}

@article{Aspelmyer2014,
	Adsurl = {https://ui.adsabs.harvard.edu/abs/2014RvMP...86.1391A},
	Archiveprefix = {arXiv},
	Author = {{Aspelmeyer}, Markus and {Kippenberg}, Tobias J. and {Marquardt}, Florian},
	Doi = {10.1103/RevModPhys.86.1391},
	Eprint = {1303.0733},
	Journal = {Reviews of Modern Physics},
	Month = oct,
	Number = {4},
	Pages = {1391-1452},
	Primaryclass = {cond-mat.mes-hall},
	Title = {{Cavity optomechanics}},
	Volume = {86},
	Year = 2014}

@article{Yang2013,
	Author = {Yang, Huan and Miao, Haixing and Lee, Da-Shin and Helou, Bassam and Chen, Yanbei},
	Doi = {10.1103/PhysRevLett.110.170401},
	Issue = {17},
	Journal = {Phys. Rev. Lett.},
	Month = {Apr},
	Numpages = {5},
	Pages = {170401},
	Publisher = {American Physical Society},
	Title = {Macroscopic Quantum Mechanics in a Classical Spacetime},
	Url = {https://link.aps.org/doi/10.1103/PhysRevLett.110.170401},
	Volume = {110},
	Year = {2013}}

@article{Helou2017,
	Author = {Helou, Bassam and Luo, Jun and Yeh, Hsien-Chi and Shao, Cheng-gang and Slagmolen, B. J. J. and McClelland, David E. and Chen, Yanbei},
	Doi = {10.1103/PhysRevD.96.044008},
	Issue = {4},
	Journal = {Phys. Rev. D},
	Month = {Aug},
	Numpages = {24},
	Pages = {044008},
	Publisher = {American Physical Society},
	Title = {Measurable signatures of quantum mechanics in a classical spacetime},
	Url = {https://link.aps.org/doi/10.1103/PhysRevD.96.044008},
	Volume = {96},
	Year = {2017}}

@article{Gan2016Optomechanical,
  title = {Optomechanical tests of a Schr\"odinger-Newton equation for gravitational quantum mechanics},
  author = {Gan, C. C. and Savage, C. M. and Scully, S. Z.},
  journal = {Phys. Rev. D},
  volume = {93},
  issue = {12},
  pages = {124049},
  numpages = {8},
  year = {2016},
  month = {Jun},
  publisher = {American Physical Society},
  doi = {10.1103/PhysRevD.93.124049},
  url = {https://link.aps.org/doi/10.1103/PhysRevD.93.124049}
}

@article{Oppenheim2023A,
  title = {A Postquantum Theory of Classical Gravity?},
  author = {Oppenheim, Jonathan},
  journal = {Phys. Rev. X},
  volume = {13},
  issue = {4},
  pages = {041040},
  numpages = {37},
  year = {2023},
  month = {Dec},
  publisher = {American Physical Society},
  doi = {10.1103/PhysRevX.13.041040},
  url = {https://link.aps.org/doi/10.1103/PhysRevX.13.041040}
}

@article{Tilloy2016,
  title = {Sourcing semiclassical gravity from spontaneously localized quantum matter},
  author = {Tilloy, Antoine and Di\'osi, Lajos},
  journal = {Phys. Rev. D},
  volume = {93},
  issue = {2},
  pages = {024026},
  numpages = {12},
  year = {2016},
  month = {Jan},
  publisher = {American Physical Society},
  doi = {10.1103/PhysRevD.93.024026},
  url = {https://link.aps.org/doi/10.1103/PhysRevD.93.024026}
}

@article{Kafri_2014,
doi = {10.1088/1367-2630/16/6/065020},
url = {https://dx.doi.org/10.1088/1367-2630/16/6/065020},
year = {2014},
month = {jun},
publisher = {IOP Publishing},
volume = {16},
number = {6},
pages = {065020},
author = {D Kafri and J M Taylor and G J Milburn},
title = {A classical channel model for gravitational decoherence},
journal = {New Journal of Physics}
}

@Article{Tilloy2024,
	title={{General quantum-classical dynamics as measurement based feedback}},
	author={Antoine Tilloy},
	journal={SciPost Phys.},
	volume={17},
	pages={083},
	year={2024},
	publisher={SciPost},
	doi={10.21468/SciPostPhys.17.3.083},
	url={https://scipost.org/10.21468/SciPostPhys.17.3.083},
}

@article{Westpal2021,
	Adsurl = {https://ui.adsabs.harvard.edu/abs/2021Natur.591..225W},
	Author = {{Westphal}, Tobias and {Hepach}, Hans and {Pfaff}, Jeremias and {Aspelmeyer}, Markus},
	Doi = {10.1038/s41586-021-03250-7},
	Journal = {\nat},
	Month = mar,
	Number = {7849},
	Pages = {225-228},
	Title = {{Measurement of gravitational coupling between millimetre-sized masses}},
	Volume = {591},
	Year = 2021}

@article{Delic2020,
	Adsurl = {https://ui.adsabs.harvard.edu/abs/2020Sci...367..892D},
	Author = {{Deli{\'c}}, Uro{\v{s}} and {Reisenbauer}, Manuel and {Dare}, Kahan and {Grass}, David and {Vuleti{\'c}}, Vladan and {Kiesel}, Nikolai and {Aspelmeyer}, Markus},
	Doi = {10.1126/science.aba3993},
	Journal = {Science},
	Month = feb,
	Number = {6480},
	Pages = {892-895},
	Title = {{Cooling of a levitated nanoparticle to the motional quantum ground state}},
	Volume = {367},
	Year = 2020}

@article{Chen_2013,
	Author = {Yanbei Chen},
	Doi = {10.1088/0953-4075/46/10/104001},
	Journal = {Journal of Physics B: Atomic, Molecular and Optical Physics},
	Month = {may},
	Number = {10},
	Pages = {104001},
	Publisher = {{IOP} Publishing},
	Title = {Macroscopic quantum mechanics: theory and experimental concepts of optomechanics},
	Url = {https://doi.org/10.1088/0953-4075/46/10/104001},
	Volume = {46},
	Year = 2013}

@article{Bassi_2017,
	Author = {Angelo Bassi and Andr{\'{e}} Gro{\ss}ardt and Hendrik Ulbricht},
	Doi = {10.1088/1361-6382/aa864f},
	Journal = {Classical and Quantum Gravity},
	Month = {sep},
	Number = {19},
	Pages = {193002},
	Publisher = {{IOP} Publishing},
	Title = {Gravitational decoherence},
	Url = {https://doi.org/10.1088/1361-6382/aa864f},
	Volume = {34},
	Year = 2017}

@article{Grossardt2016,
	Author = {Gro\ss{}ardt, Andr\'e and Bateman, James and Ulbricht, Hendrik and Bassi, Angelo},
	Doi = {10.1103/PhysRevD.93.096003},
	Issue = {9},
	Journal = {Phys. Rev. D},
	Month = {May},
	Numpages = {6},
	Pages = {096003},
	Publisher = {American Physical Society},
	Title = {Optomechanical test of the Schr\"odinger-Newton equation},
	Url = {https://link.aps.org/doi/10.1103/PhysRevD.93.096003},
	Volume = {93},
	Year = {2016}}

@article{Carney_2019,
	Author = {Daniel Carney and Philip C E Stamp and Jacob M Taylor},
	Doi = {10.1088/1361-6382/aaf9ca},
	Journal = {Classical and Quantum Gravity},
	Month = {jan},
	Number = {3},
	Pages = {034001},
	Publisher = {{IOP} Publishing},
	Title = {Tabletop experiments for quantum gravity: a user's manual},
	Url = {https://doi.org/10.1088/1361-6382/aaf9ca},
	Volume = {36},
	Year = 2019}

@book{Braginsky1995,
	Adsurl = {https://ui.adsabs.harvard.edu/abs/1995qume.book.....B},
	Author = {{Braginsky}, Vladimir B. and {Khalili}, Farid Ya and {Thorne}, Kip S.},
	Title = {{Quantum Measurement}},
	Year = 1995}

@article{Rossi2019Observing,
  title = {Observing and Verifying the Quantum Trajectory of a Mechanical Resonator},
  author = {Rossi, Massimiliano and Mason, David and Chen, Junxin and Schliesser, Albert},
  journal = {Phys. Rev. Lett.},
  volume = {123},
  issue = {16},
  pages = {163601},
  numpages = {6},
  year = {2019},
  month = {Oct},
  publisher = {American Physical Society},
  doi = {10.1103/PhysRevLett.123.163601},
  url = {https://link.aps.org/doi/10.1103/PhysRevLett.123.163601}
}

@book{Wiener1964Extrapolation, 
author = {Wiener, Norbert}, 
title = {Extrapolation, Interpolation, and Smoothing of Stationary Time Series},
 year = {1964}, 
 isbn = {0262730057}, 
 publisher = {The MIT Press} ,
 url = {https://ieeexplore.ieee.org/servlet/opac?bknumber=6267356}
 }

@article{Carney2022,
	Author = {Carney, Daniel},
	Doi = {10.1103/PhysRevD.105.024029},
	Issue = {2},
	Journal = {Phys. Rev. D},
	Month = {Jan},
	Numpages = {17},
	Pages = {024029},
	Publisher = {American Physical Society},
	Title = {Newton, entanglement, and the graviton},
	Url = {https://link.aps.org/doi/10.1103/PhysRevD.105.024029},
	Volume = {105},
	Year = {2022}}

@article{Miao2020,
	Author = {Miao, Haixing and Martynov, Denis and Yang, Huan and Datta, Animesh},
	Doi = {10.1103/PhysRevA.101.063804},
	Issue = {6},
	Journal = {Phys. Rev. A},
	Month = {Jun},
	Numpages = {7},
	Pages = {063804},
	Publisher = {American Physical Society},
	Title = {Quantum correlations of light mediated by gravity},
	Url = {https://link.aps.org/doi/10.1103/PhysRevA.101.063804},
	Volume = {101},
	Year = {2020}}

@article{Cripe2019,
	Adsurl = {https://ui.adsabs.harvard.edu/abs/2019Natur.568..364C},
	Author = {{Cripe}, Jonathan and {Aggarwal}, Nancy and {Lanza}, Robert and {Libson}, Adam and {Singh}, Robinjeet and {Heu}, Paula and {Follman}, David and {Cole}, Garrett D. and {Mavalvala}, Nergis and {Corbitt}, Thomas},
	Doi = {10.1038/s41586-019-1051-4},
	Journal = {\nat},
	Month = mar,
	Number = {7752},
	Pages = {364-367},
	Title = {{Measurement of quantum back action in the audio band at room temperature}},
	Volume = {568},
	Year = 2019}

@article{Datta_2021,
	Author = {Animesh Datta and Haixing Miao},
	Doi = {10.1088/2058-9565/ac1adf},
	Journal = {Quantum Science and Technology},
	Month = {aug},
	Number = {4},
	Pages = {045014},
	Publisher = {{IOP} Publishing},
	Title = {Signatures of the quantum nature of gravity in the differential motion of two masses},
	Url = {https://doi.org/10.1088/2058-9565/ac1adf},
	Volume = {6},
	Year = 2021}

@article{Snowmass2022,
	Adsurl = {https://ui.adsabs.harvard.edu/abs/2022arXiv220311846C},
	Archiveprefix = {arXiv},
	Author = {{Carney}, Daniel and {Chen}, Yanbei and {Geraci}, Andrew and {M{\"u}ller}, Holger and {Panda}, Cristian D. and {Stamp}, Philip C.~E. and {Taylor}, Jacob M.},
	Eid = {arXiv:2203.11846},
	Eprint = {2203.11846},
	Journal = {arXiv e-prints},
	Month = mar,
	Pages = {arXiv:2203.11846},
	Primaryclass = {gr-qc},
	Title = {{Snowmass 2021 White Paper: Tabletop experiments for infrared quantum gravity}},
	Year = 2022}

@article{Matsumoto2020,
	Author = {Cata\~no-Lopez, Seth B. and Santiago-Condori, Jordy G. and Edamatsu, Keiichi and Matsumoto, Nobuyuki},
	Doi = {10.1103/PhysRevLett.124.221102},
	Issue = {22},
	Journal = {Phys. Rev. Lett.},
	Month = {Jun},
	Numpages = {6},
	Pages = {221102},
	Publisher = {American Physical Society},
	Title = {High-$Q$ Milligram-Scale Monolithic Pendulum for Quantum-Limited Gravity Measurements},
	Url = {https://link.aps.org/doi/10.1103/PhysRevLett.124.221102},
	Volume = {124},
	Year = {2020}}

@article{Carlip_2008,
	Author = {S Carlip},
	Doi = {10.1088/0264-9381/25/15/154010},
	Journal = {Classical and Quantum Gravity},
	Month = {jul},
	Number = {15},
	Pages = {154010},
	Publisher = {{IOP} Publishing},
	Title = {Is quantum gravity necessary?},
	Url = {https://doi.org/10.1088/0264-9381/25/15/154010},
	Volume = {25},
	Year = 2008}

@article{SalzmanCarlip2006,
	Author = {Peter Jay Salzman and Steven Carlip},
	Archiveprefix = {arXiv},
	Eprint = {gr-qc/0606120},
	Journal = {arXiv preprint gr-qc/0606120},
	Primaryclass = {gr-qc},
	Title = {A possible experimental test of quantized gravity},
	Url = {https://arxiv.org/abs/gr-qc/0606120},
	Year = {2006}}

@article{Diosi1989,
	Author = {Di\'osi, L.},
	Doi = {10.1103/PhysRevA.40.1165},
	Issue = {3},
	Journal = {Phys. Rev. A},
	Month = {Aug},
	Numpages = {0},
	Pages = {1165--1174},
	Publisher = {American Physical Society},
	Title = {Models for universal reduction of macroscopic quantum fluctuations},
	Url = {https://link.aps.org/doi/10.1103/PhysRevA.40.1165},
	Volume = {40},
	Year = {1989}}

@article{Diosi1998,
	Author = {Di\'osi, Lajos and Halliwell, Jonathan J.},
	Doi = {10.1103/PhysRevLett.81.2846},
	Issue = {14},
	Journal = {Phys. Rev. Lett.},
	Month = {Oct},
	Numpages = {0},
	Pages = {2846--2849},
	Publisher = {American Physical Society},
	Title = {Coupling Classical and Quantum Variables using Continuous Quantum Measurement Theory},
	Url = {https://link.aps.org/doi/10.1103/PhysRevLett.81.2846},
	Volume = {81},
	Year = {1998}}

@article{Penrose1996,
	Author = {Penrose, Roger},
	Da = {1996/05/01},
	Doi = {10.1007/BF02105068},
	Id = {Penrose1996},
	Isbn = {1572-9532},
	Journal = {General Relativity and Gravitation},
	Number = {5},
	Pages = {581--600},
	Title = {On Gravity's role in Quantum State Reduction},
	Ty = {JOUR},
	Url = {https://doi.org/10.1007/BF02105068},
	Volume = {28},
	Year = {1996}}

@article{Liu2023,
  title = {Semiclassical gravity phenomenology under the causal-conditional quantum measurement prescription},
  author = {Liu, Yubao and Miao, Haixing and Chen, Yanbei and Ma, Yiqiu},
  journal = {Phys. Rev. D},
  volume = {107},
  issue = {2},
  pages = {024004},
  numpages = {18},
  year = {2023},
  month = {Jan},
  publisher = {American Physical Society},
  doi = {10.1103/PhysRevD.107.024004},
  url = {https://link.aps.org/doi/10.1103/PhysRevD.107.024004}
}

@article{Liu2024,
  title = {Semiclassical gravity phenomenology under the causal-conditional quantum measurement prescription. II. Heisenberg picture and apparent optical entanglement},
  author = {Liu, Yubao and Zhong, Wenjie and Chen, Yanbei and Ma, Yiqiu},
  journal = {Phys. Rev. D},
  volume = {111},
  issue = {6},
  pages = {062004},
  numpages = {26},
  year = {2025},
  month = {Mar},
  publisher = {American Physical Society},
  doi = {10.1103/PhysRevD.111.062004},
  url = {https://link.aps.org/doi/10.1103/PhysRevD.111.062004}
}

@article{Miki2025,
      title={The Role of Quantum Measurements when Testing the Quantum Nature of Gravity}, 
      author={Daisuke Miki and Youka Kaku and Yubao Liu and Yiqiu Ma and Yanbei Chen},
       journal =  {arXiv: 2503.11882},
       year={2025},
       url={https://arxiv.org/abs/2503.11882}, 
}

@article{Giulini2014,
	Author = {Giulini, Domenico and Gro{\ss}ardt, Andr{\'e}},
	Journal = {New Journal of Physics},
	Number = {7},
	Pages = {075005},
	Publisher = {IOP Publishing},
	Title = {Centre-of-mass motion in multi-particle Schr{\"o}dinger--Newton dynamics},
	Volume = {16},
	Year = {2014}}

@article{Giulini2011,
	Author = {Giulini, Domenico and Gro{\ss}ardt, Andr{\'e}},
	Journal = {Classical and Quantum Gravity},
	Number = {19},
	Pages = {195026},
	Publisher = {IOP Publishing},
	Title = {Gravitationally induced inhibitions of dispersion according to the Schr{\"o}dinger--Newton equation},
	Volume = {28},
	Year = {2011}}

@article{Bose2025,
  title = {Massive quantum systems as interfaces of quantum mechanics and gravity},
  author = {Bose, Sougato and Fuentes, Ivette and Geraci, Andrew A. and Khan, Saba Mehsar and Qvarfort, Sofia and Rademacher, Markus and Rashid, Muddassar and Toro\ifmmode \check{s}\else \v{s}\fi{}, Marko and Ulbricht, Hendrik and Wanjura, Clara C.},
  journal = {Rev. Mod. Phys.},
  volume = {97},
  issue = {1},
  pages = {015003},
  numpages = {71},
  year = {2025},
  month = {Feb},
  publisher = {American Physical Society},
  doi = {10.1103/RevModPhys.97.015003},
  url = {https://link.aps.org/doi/10.1103/RevModPhys.97.015003}
}

@article{Yan2025,
  title = {First result for testing semiclassical gravity effect with a torsion balance},
  author = {Yan, Tianliang and Prokhorov, Leonid and Smetana, Jiri and Boyer, Vincent and Martynov, Denis and Liu, Yubao and Ma, Yiqiu and Miao, Haixing},
  journal = {Phys. Rev. D},
  volume = {111},
  issue = {8},
  pages = {082007},
  numpages = {17},
  year = {2025},
  month = {Apr},
  publisher = {American Physical Society},
  doi = {10.1103/PhysRevD.111.082007},
  url = {https://link.aps.org/doi/10.1103/PhysRevD.111.082007}
}

@Article{Smetana2024,
AUTHOR = {Smetana, Jiri and Yan, Tianliang and Boyer, Vincent and Martynov, Denis},
TITLE = {A High-Finesse Suspended Interferometric Sensor for Macroscopic Quantum Mechanics with Femtometre Sensitivity},
JOURNAL = {Sensors},
VOLUME = {24},
YEAR = {2024},
NUMBER = {7},
ARTICLE-NUMBER = {2375},
URL = {https://www.mdpi.com/1424-8220/24/7/2375},
PubMedID = {38610586},
ISSN = {1424-8220},
DOI = {10.3390/s24072375}
}

@article{Zhong2025,
  title = {Distinguishing quantum and classical gravity via nonstationary test mass dynamics},
  author = {Zhong, Wenjie and Liu, Yubao and Ma, Yiqiu},
  journal = {Phys. Rev. D},
  volume = {112},
  issue = {4},
  pages = {044060},
  numpages = {21},
  year = {2025},
  month = {Aug},
  publisher = {American Physical Society},
  doi = {10.1103/nl32-g2r4},
  url = {https://link.aps.org/doi/10.1103/nl32-g2r4}
}

@article{BialynickiBirula1976,
  title = {Nonlinear Wave Mechanics},
  author = {Bialynicki-Birula, I. and Mycielski, J.},
  journal = {Annals of Physics},
  volume = {100},
  number = {1-2},
  pages = {62--93},
  year = {1976},
  doi = {10.1016/0003-4916(76)90057-9}
}

@article{Weinberg1989,
  title = {Testing Quantum Mechanics},
  author = {Weinberg, Steven},
  journal = {Annals of Physics},
  volume = {194},
  number = {2},
  pages = {336--386},
  year = {1989},
  doi = {10.1016/0003-4916(89)90276-5}
}

@article{Gisin1990,
  title = {Weinberg's Non-Linear Quantum Mechanics and Supraluminal Communications},
  author = {Gisin, Nicolas},
  journal = {Physics Letters A},
  volume = {143},
  number = {1-2},
  pages = {1--2},
  year = {1990},
  doi = {10.1016/0375-9601(90)90786-N}
}

@article{Polchinski1991,
  title = {Weinberg's Nonlinear Quantum Mechanics and the Einstein-Podolsky-Rosen Paradox},
  author = {Polchinski, Joseph},
  journal = {Physical Review Letters},
  volume = {66},
  number = {4},
  pages = {397--400},
  year = {1991},
  doi = {10.1103/PhysRevLett.66.397}
}

\end{document}